\documentclass[11pt,a4paper]{article}

\pdfoutput=1
\usepackage{jcappub}
\usepackage{graphicx}% Include figure files
\usepackage{dcolumn}% Align table columns on decimal point
\usepackage{bm}% bold math
\usepackage{color}
\usepackage{calrsfs}
\usepackage{hyperref}
\usepackage{multirow}
\usepackage{float}
\usepackage{soul}  
\usepackage{amsmath}
\input epsf

\newcommand{\la}{\,\rlap{\raise 0.5ex\hbox{$<$}}{\lower 1.0ex\hbox{$\sim$}}\,}
\newcommand{\ga}{\,\rlap{\raise 0.5ex\hbox{$>$}}{\lower 1.0ex\hbox{$\sim$}}\,}

\newcommand{\be}{\begin{equation}}
\newcommand{\e}{\end{equation}}
\def\bea{\begin{eqnarray}}
\def\eea{\end{eqnarray}}
\newcommand{\bear}{\begin{eqnarray}}
\newcommand{\ear}{\end{eqnarray}}

\newcommand{\pin}{{\sc PINOCCHIO}}

\begin{document}

\title{A blind method to recover the mask of a deep galaxy survey}

%%%%%%%%%%%%%% AUTHORS %%%%%%%%%%%%
\author[a,b,c]{Pierluigi Monaco,} 
\author[d,e]{Enea Di Dio,}
\author[b,c]{Emiliano Sefusatti}

\affiliation[a]{Dipartimento di Fisica, Sezione di Astronomia, via G.B. Tiepolo 11, I-34143 Trieste, Italy}
\affiliation[b]{INAF - Astronomical Observatory of Trieste, via G.B. Tiepolo 11, I-34143 Trieste, Italy}
\affiliation[c]{INFN -- National Institute for Nuclear Physics, Via Valerio 2, I-34127 Trieste, Italy}
\affiliation[d]{Lawrence Berkeley National Laboratory,\\ 1 Cyclotron Road, Berkeley, CA 93720, USA}
\affiliation[e]{Berkeley Center for Cosmological Physics,\\ University of California, Berkeley, CA 94720, USA}

\emailAdd{pierluigi.monaco@inaf.it}
\emailAdd{enea.didio@berkeley.edu}
\emailAdd{emiliano.sefusatti@inaf.it}

%%%%%%%%%%%%%%%%% ABSTRACT %%%%%%%%%%%%%%%%%%
\abstract{
We present a blind method to determine the properties of a foreground
contamination, given by a visibility mask, that affects a deep galaxy
survey. Angular cross correlations of density fields in different
redshift bins are expected to vanish (apart from a contribution due to
lensing), but are sensitive to the presence of a foreground that
modulates the flux limit across the sky. After formalizing the
expected effect of a foreground mask on the measured galaxy density,
under a linear, luminosity-dependent bias model for galaxies, we
construct two estimators, based on the average or square average
galaxy density in a given sky pixel along the line of sight, that
single out the mask contribution if a sufficient number of independent
redshift bins is available. These estimators are combined to give a
reconstruction of the mask. We use Milky-Way reddening as a prototype
for the mask. Using a set of 20 large mock catalogs  covering
 $1/4$-th of the sky and number-matched to $H\alpha$ emitters to
mimic an Euclid-like sample, we demonstrate that our method can
reconstruct the mask and its angular clustering at scales $\ell
\la100$, beyond which the cosmological signal becomes dominant. The
uncertainty of this reconstruction is quantified to be $1/3$-rd of
the sample variance of the signal. Such a reconstruction requires
knowledge of the average and square average of the mask,
but we show that it is
possible to recover this information either from external models or
internally from the data. It also relies on knowledge of how the
impact of the foreground changes with redshift (due to the extinction
curve in our case), but this can be tightly constrained by
cross correlations of different redshift bins. The strong points of
this blind reconstruction technique lies in the ability to find
``unknown unknowns'' that affect a survey, and in the facility to
quantify, using sets of mock catalogs, how its uncertainty propagates
to clustering measurements. }
  
\maketitle

%%%%%%%%%%%%%%%%%%%% INTRODUCTION %%%%%%%%%%%%%%%%%
\section{Introduction}
\label{sec:introduction}

The next generation of galaxy surveys, like
Euclid~\cite{laureijs2011}, DESI~\cite{frieman2013},
WFIRST~\cite{green2012}, LSST~\cite{abell2009},
eBoss~\cite{dawson2016} and the SKA~\cite{Maartens:2015mra} surveys,
will map large fractions of the sky, going deep, well beyond redshift
one, and collecting data for billions of galaxies. They will provide
measurements of galaxy clustering to unprecedented sub-percent
precision, providing tight constraints to parameters like the equation
of state of dark energy and its possible redshift evolution, that may
discriminate a pure cosmological constant from a dynamical component.
With statistical errors beaten down, the error budget will be
dominated by systematics. A class of these systematics is connected
with foregrounds on the sky, internal to the Milky Way or related to
the nearest galaxies, that modulate the effective depth of the survey.
We can broadly divide such foregrounds into those that increase the
background noise, like zodiacal light, and those that decrease the
signal, like Milky Way extinction. These foregrounds, if not properly
corrected for, create large-scale structure on degree scales, that is
then projected to the observed redshift to create fake structures at
the $\sim$Gpc comoving scales that lie before (in wave number) or at
the peak of the galaxy power spectrum.

A very similar effect can be induced by more instrumental issues, like
errors in the zero-point calibration of photometry, straylight within
the detector, image persistence in the CCD, modulations of survey
exposure time or decay of the instrument performance along its
lifetime, fiber collision or overlapping spectra in slitless
spectroscopy. The impact of these features couples with the survey
strategy to determine the effective flux limit of the survey across
the sky.

Every team involved in the preparation of the future surveys mentioned
above is working hard to model and subtract out these effects. Most of
the approaches proposed so far are based either on prior knowledge of
the contaminant, or on removal of the most contaminated modes, with
some information loss. For instance, removal of contamination due to
fiber collision can be done, with some loss of information, by
removing purely angular modes in the anisotropic power spectrum
\cite{Burden2017,Pinol2017}. Experience with previous surveys teaches
us that real data are often affected by ``unknown unknowns'', i.e. sources
of contamination that were not accounted for in the preliminary
analysis. An example is given by the SDSS-III BOSS survey
\cite{Ross2011}, where stars of the same magnitude as the observed
galaxies were decreasing the effective area of the survey, creating
spurious power when this decrease was not corrected for. The
contamination was found to be strong at scales just beyond the first
Baryonic Acoustic Oscillation (BAO), and its cause was found by
correlating the galaxy density with a list of potential causes.
Correction of systematics in \cite{Ho2012,Ross2012} was based on
template subtraction, assuming that each potential foreground is
contributing linearily to the observed density contrast; the
corresponding linear coefficients were found by cross-correlating the
signal with the template. This method of template subtraction, coupled
with mode deprojection, was further developed by
\cite{Kalus2016,Kalus2019}. Besides correlating the observed density
with known sources, an often used approach is that of computing cross
correlations of very different surveys that are subject to different
systematics; in this case it is in principle possible to minimise the
contamination. This technique has been used to constrain primordial
non-Gaussianities
\cite{Camera2013,Giannantonio2014a,Giannantonio2014b}. A caveat to
this approach is that the Galaxy is a source of contamination at all
wavelengths, so it is not obvious to understand at what level
foregrounds at different wavelengths are really uncorrelated. 21 cm
intensity mapping analysis exploits the idea that foregrounds are
modulated on very different scales, and this makes it possible to
perform a blind separation of signal and foreground \cite{Alonso2015},
at the cost of removing large-scale cosmological modes that lie in the
same domain as the contaminants. Besides, the analysis of the Cosmic
Microwave Background (CMB) can exploit, with component separation
techniques, the fact that different components have different Specral
Energy Distributions (SEDs) \cite{Planck2018}, reaching a much higher
precision level in foreground subtraction. One could think to extend
this principle to galaxy surveys using redshift in place of
wavelength: foregrounds could be fitted based on of how their impact
on the observed density changes with redshift, see \cite{Jasche2017,Porqueres2018}
for an example in this direction.

Several foregrounds can be readily subtracted once images are acquired
and well calibrated; for instance, the precise origin of the
background noise in an image (say zodiacal light, extragalactic
background, or straylight) should be known in advance to simulate a
survey and assess its completeness, but when images have been taken
the noise level is directly measured. The case of Milky Way extinction
is different, as the extragalactic light is absorbed by dust grains
and reprocessed in the mid and far infrared (FIR), and there is no internal
way to know the fraction of absorbed light. Extinction is typically
subtracted using models based on observed FIR emission maps of the
Galaxy. In Schlegel, Finkbeiner \& Davis \cite{SFD} (hereafter SFD)
the authors used COBE and IRAS data to create a map of FIR emission,
then transformed this into a reddening map by calibrating it with reddening
values measured for a set of galaxies.
More recently, the Planck Collaboration issued two extinction or
reddening maps, based on the same principles. The first map was issued
with the 2013 results \cite{P13} (hereafter P13): they used 353, 545
and 857 GHz maps, together with IRAS 100 $\mu$m map, and fitted the
SED of each pixel with a modified black body. They then used the
radiance as a quantity supposedly well correlated with extinction, and
calibrated the coefficient of this linear correlation with reddening
measured for a set of 53,399 SDSS quasars. Care was taken in
minimizing the impact of Cosmic Infrared Background (CIB) fluctuations
on reddening estimates. In 2015 a different map was issued, based on
Planck intermediate results \cite{P15} (hereafter P15): they added to
the previous maps the IRAS 60 $\mu$m and the WISE 12 $\mu$m bands, and
fitted pixel-by-pixel SEDS with a physical dust model by Draine \& Li
\cite{Draine2007}, where extinction $A_V$ is a parameter of the model.
However, a check with 272,366 SDSS quasars showed that extinction was
overestimated by a factor of 2.4, so the extinction maps were
corrected by that heuristic factor. We will show in this paper that
the level of agreement of these three reddening or extinction maps is
insufficient to guarantee the required accuracy in foreground
subtraction.

These foregrounds are expected to be dominant at very large scales. In
fact, at scales beyond the peak of the power spectrum the clustering
signal is expected to be weak, while foregrounds, being driven by
local astrophysical entities that surround the Earth, are clustered on
very large angular scales; the same is true for instrumental and
survey issues (with remarkable exceptions of fiber collision or
spectra overlap that will be discussed in the conclusions). While
their impact at the very important BAO scales is typically under
control, measuring clustering at the largest scales will be limited by
foreground subtraction. There are several science cases that require
accurate measurements at the largest scale. In the standard scenario,
these scales have been frozen once they exited the Hubble horizon
during the inflationary epoch, and they are the neatest fossil to
understand the physical interactions at the inflationary energy scale.
These interactions may have generated some primordial non-gaussianity,
whose amplitude is usually parametrized by $f_{\rm nl}$. Planck has
constrained this parameter to be $f_{\rm nl}=0.8\pm5.0$
\cite{PlanckNL}. While the search of generic primordial
non-gaussianity requires an analysis beyond the 2-point statistics,
local non-gaussianity generates a well-defined scale-dependent galaxy
bias~\cite{Dalal:2007cu} which is enhanced on large scales. This can
be exploited to achieve tighter constraints than Planck's, especially
using multiple tracers~\cite{Seljak:2008xr} in future redshift
surveys. Moreover, at the largest scales the simple Newtonian
approximation may start to fail; in the last decade several efforts
have been devoted to describing galaxy clustering in a gauge-invariant
relativistic
framework~\cite{Yoo:2009au,Bonvin:2011bg,Challinor:2011bk} to meet the
experimental accuracy of coming surveys. These effects are expected to
be relevant at the horizon scale and their amplitude and detectability
are comparable with $f_{\rm nl} \sim
\mathcal{O}(1)$~\cite{Alonso:2015sfa}.

Proper subtraction of foregrounds is only part of the problem, the
uncertainty on foreground removal should be quantified in terms of its
contribution to the covariance matrix of clustering measurements, and
propagated to parameter estimation. In a recent paper, Colavincenzo et
al. \cite{Colavincenzo2017} (hereafter C17), the authors quantified,
with an idealized foreground model, the impact of the uncertainty of
foreground subtraction on the covariance matrix of the power spectrum
of dark matter halos, showing that, although cosmological clustering
and foreground are assumed to be uncorrelated, the full covariance matrix is
not simply the sum of the cosmological one and that induced by the
foreground, because of a rather long list of mixed terms that can be quite
significant.

To determine the full covariance matrix it is necessary to know not
only the error on foreground subtraction but also its correlations on
the sky. Foregrounds mainly impact clustering measurements through a
modulation of survey depth. We will loosely call ``visibility mask'',
or simply ``mask'', a function of the sky position $\boldsymbol\theta$
that is responsible for the modulation of survey depth, though its
specific impact may in general depend on the target redshift. Throughout the
paper we will use galactic extinction (more specifically, $E(B-V)$
reddening) as a prototype mask. The starting idea is that (see e.g.
C17) to first order a foreground mask ${\cal M}$ acts on the survey in
such a way that the observed density contrast $\delta_{\rm o}$ is
related to the true galaxy density contrast $\delta_{\rm g}$ as
$1+\delta_{\rm o}=(1-{\cal M})(1+\delta_{\rm g})$ (see equation (2.19)
of C17). The two-point correlation function of the density is then:

\begin{equation}
\langle \delta_{\rm o1}\delta_{\rm o2}\rangle = 
\langle \delta_{\rm g1}\delta_{\rm g2}\rangle +
\langle {\cal M}_1  {\cal M}_2\rangle +
\langle \delta_{\rm g1}\delta_{\rm g2}\rangle \langle {\cal M}_1 {\cal M}_2\rangle\, .
\label{eq:start}
\end{equation}

\noindent
Then, if we measure a cosmic correlation that is expected to be
vanishing, $\langle \delta_{\rm g1}\delta_{\rm g2}\rangle=0$, like the
angular cross-correlation of two different and distant redshift bins,
any significant signal could be interpreted as the angular correlation
of the foreground, and used to reconstruct it. This will provide
accurate results as long as the survey is deep enough to have a large
number of independent pairs of redshift bins.

In this paper we build on this idea, presenting a blind method to
recover the mask to which a cosmological catalog is subject. The
method will be applied to a set of mock galaxy catalogs on the past
light cone covering $1/4$ of the sky, generated with the {\pin}
approximate method \cite{Monaco2013,Munari2017}, and calibrated to
generate an Euclid-like survey of $H\alpha$ emitters
\citep{Pozzetti2016} in a wide redshift range. The reconstruction of
the mask will be based on estimators of the average density contrast
along the line of sight; the inversion of the relation between mask
and estimator requires knowledge of the first two moments of the mask,
and we will show how this information can be achieved with sufficient
accuracy from an internal analysis of the catalog. We will check that
the cross correlation induced by light-cone effects, like
magnification bias due to lensing, does not affect the scale where our
prototypical mask is dominant and can be recovered. We will further
show to what level this reconstruction can be used to test different
maps of galactic extinction, in the case this is the only present
foreground.

The paper is organized as follows. Section~\ref{sec:impact} gives a
formalization of the problem, defining the observed density contrast
and its dependence on the flux limit of the survey, and expanding the
equations in the case of galactic extinction as foreground.
Section~\ref{sec:mocks} describes the simulated catalogs of dark
matter halos in the past light-cone and the abundance-matching
connection of halos with galaxies. Section~\ref{sec:reconstructions}
describes the procedure adopted to obtain a ``best reconstruction''
for the mask, starting from estimators of galaxy clustering that are
sensitive only to the mask term, and under the hypothesis that the
first two moments of the mask are known. Section~\ref{sec:cl}
discusses how the angular clustering of the cosmological, unmasked
mocks can be recovered using the mask best reconstruction, and the
crucial role of cross correlations for the calibration of the mask.
Section~\ref{sec:moments} discusses how the first two moments of the
mask can be estimated from the available data, and tests the
performance of a best reconstruction applied with no external
information and calibrated on cross-correlations.
Section~\ref{sec:testing} shows to what extent an application of this
procedure can be used to test reddening maps and recover the
extinction curve. Section~\ref{sec:conclusions} presents a final
discussion and the conclusions.

\section{The impact of a visibility mask on clustering measurements}
\label{sec:impact}

\subsection{The galaxy density contrast}

A density measurement is based on counting galaxies that are observed
in specified volumes. The condition for a galaxy to be observed is
typically expressed as a flux limit threshold. In realistic cases, the
probability of a galaxy being observed will be a smooth function of
flux; this will be quantified by a completeness function, obtained
through detailed simulations of the observational setup. Knowledge of
the completeness function allows one to recast the observational
condition into a strict flux limit threshold, that at a certain
redshift $z$ can be recast into a luminosity threshold $L_0(z)$ by
assuming a fiducial cosmology.

We call $\Phi(L|z)$ the galaxy luminosity function at redshift $z$,
and $\phi_{\rm local}(L|\mathbf{x})$ the {\em local} luminosity
function in a volume centred on the comoving position $\mathbf{x}$,
lying at the redshift $z$: $|\mathbf{x}|=d_c(z)$, where $d_c(z)$ is
the comoving distance from the observer. The relation between $\Phi$
and $\phi_{\rm local}$ is given by an average over the shell of the
survey volume within a redshift bin $\Delta z$ around $z$. We denote
such an average by $\langle\cdots\rangle$, so that:
\be
 \Phi(L|z) = \langle\phi_{\rm local}(L|{\mathbf x})\rangle\, .
\label{eq:phi}
\e
We stress that $\langle\cdots\rangle$ does not denote an ensemble
average in this paper. The integrals in luminosity, from $L_0$ to
$\infty$, of $\phi_{\rm local}$ and $\Phi$ give, respectively, the
local galaxy density $n_g(\mathbf{x}|L_0)$ and the average galaxy density
$\langle n_{\rm g} \rangle(z|L_0)$:
\be
  n_g(\mathbf{x}|L_0) = \int_{L_0(z)}^\infty \phi_{\rm local}(L|{\mathbf x}) dL\, , \ \ \ \ \ \ \ \ \ 
  \langle n_{\rm g}\rangle  (z|L_0) = \int_{L_0(z)}^\infty \Phi(L|z) dL\, .
  \label{eq:densities}
\e

Under the assumption that the shape of the luminosity function is
independent of environment, $\phi_{\rm local}(L|\mathbf{x})$ differs
from $\Phi(L|z)$ only in the normalization:

\be
  \phi_{\rm local}(L|\mathbf{x}) = [1+\delta_\phi(\mathbf{x})] \Phi(L|z)\, .
\label{eq:philocal}
  \e

\noindent
We define the galaxy density contrast in the usual way:
\be
\delta_{\rm g} = \frac{n_g -\langle n_{\rm g}\rangle}{\langle n_{\rm g}\rangle}\, .
\e
We are using here two different symbols for $\delta_\phi$ and
$\delta_{\rm g}$ because the two quantities have a different relation
with the underlying matter density contrast, due to the fact that
galaxy bias in general depends on galaxy luminosity. We will restrict
ourself here to a simple linear, luminosity-dependent bias model to
relate the galaxy density contrast $\delta_{\rm g}$ to the matter
density contrast $\delta$ in the same volume used to count galaxies:
\be
  \delta_{\rm g} (\mathbf{x}|L_0) = b_1(L_0,z) \delta(\mathbf{x}) \, , \ \ \ \ \ \ \ \ \ 
  \delta_\phi  (\mathbf{x}|L) = \beta_1(L,z) \delta(\mathbf{x})
\label{eq:bias}
\e
where $b_1(L_0,z)$ is the galaxy linear bias and
\be
b_1(L_0,z) = \frac{1}{\langle n_{\rm g}(z|L_0)\rangle}
    \int_{L_0(z)}^\infty \beta_1(L,z)\Phi(L|z)dL \label{eq:b_and_beta}\,.
\e

\noindent
Getting back to equation~\ref{eq:philocal}, we note that a
luminosity-dependent bias implies that $\delta_\phi$ gets a dependence
on luminosity (equation~\ref{eq:bias}), and so the shape of the local
luminosity function is not strictly universal. We will neglect this
second-order effect in this paper.

\subsection{Observed density contrast in presence of a foreground}

A foreground contamination will have the effect of modulating the flux
limit on the sky. We denote here the sky position with the
two-dimensional vector $\boldsymbol\theta$; this is related to the
comoving position $\mathbf{x}$ as its angular part in a spherical
coordinate system, the radius $r=|\mathbf{x}|$ being given by
$d_c(z)$. We call $L_{\rm lim}(\boldsymbol\theta,z)$ the flux limit in
a given angular position $\boldsymbol\theta$ of the sky, pointing to
the position $\mathbf{x}$. In the following, we will not explicitly
write the $z$-dependence of the various quantities. The observed
number density will thus be:
\be
n_{\rm o}(\mathbf{x}) = \int_{L_{\rm lim}(\boldsymbol\theta)}^\infty \phi_{\rm local} (L|\mathbf{x})dL\, .
\label{eq:nobs}
\e
We can write the luminosity threshold as:
\be
L_{\rm lim}(\boldsymbol\theta) = L_0 + \delta L(\boldsymbol\theta)\, .
\label{eq:deltaL}
\e
In general, the perturbation $\delta L(\boldsymbol\theta)$ will {\em
  not} average to zero over the survey. This is easy to see in the
case of galactic extinction: $L_0$ will be the luminosity limit
without extinction, and $\delta L$ will always be positive, as
extinction acts in increasing the limiting flux and luminosity.

For small perturbations of the luminosity threshold $\delta L$, one
can Taylor-expand the observed density. It is convenient to define the
quantity
\be
\epsilon \equiv \delta L/L_0 \label{eq:epsilon} 
\e
and two redshift-dependent scaling functions related to the luminosity
function and its $L$-derivative:
\be
S_A(z) \equiv \frac {\langle n_{\rm g}\rangle(z)} {\Phi(L_0|z)L_0(z)}\, ,
\label{eq:SAz}
\e
 \be
 S_B(z) \equiv \frac{d\Phi}{dL}(L_0|z) \frac{L_0(z)}{2\Phi(L_0|z)} + \frac{1}{2}\, .
\label{eq:SBz}
\e
Both scaling functions depend only on the shape of the luminosity
function beyond or at $L_0$. The convenience of adding $1/2$ to $S_B$
will be clear later. It is worth noticing that $2S_B-1 = d
\ln\Phi/d\ln L $ evaluated at $L=L_0$, and if $\Phi$ is modeled as a Schechter
function with slope $\alpha$ (including its negative sign) and
characteristic luminosity $L_\star$, then $S_B = (\alpha - L_0/L_\star
+ 1)/2$.

The observed density can be thus written as:
\be
  \frac{n_{\rm o}}{\Phi L_0} =   \frac{n_{\rm g}}{\Phi L_0} -
  (1+\beta_1\delta)\, \epsilon - \left(S_B-\frac{1}{2}\,\right)
  \left[1+\left(\beta_1+\beta_1' \frac{\Phi}{\Phi'}\right)\delta\right]\, \epsilon^2  + O(\epsilon^3)
\, .
\label{eq:expanded}
\e
 It is convenient to note that $n_{\rm g}/\Phi L_0 =
 S_A(1+b_1\delta)$. In this equation the prime denotes a derivative
 with respect to $L$, and again all $L$-dependent functions are
 evaluated at $L_0$.

To compute $\langle n_{\rm o}\rangle$ we can assume no correlation between
$\delta L(\boldsymbol\theta)$ (and thus $\epsilon(\boldsymbol\theta)$)
and $\delta(\mathbf{x})$. This is true, in an ensemble-average sense,
whenever the foreground is independent of the cosmological signal.
This covers most foregrounds, with some remarkable exceptions, the
obvious cases being density-dependent biases like fiber collision or
spectra overlapping in slitless spectroscopy. Moreover, a reddening
map obtained from FIR observations will be contaminated by the Cosmic
Infrared Background, that gets a contribution by the same surveyed
galaxies (see discussion in \cite{P13}), so if extinction is corrected
for using such a map the residual foreground will be correlated with
the cosmological signal; we will further discuss this point in
Section~\ref{sec:conclusions}. We should notice, however, that even in the case of no correlation between $\epsilon$ and $\delta$, a given volume average does not guarantee
that terms like $\langle \epsilon \delta \rangle$ go to zero exactly.
We do not attempt here an estimation of the magnitude of these
residual terms. Their impact will be implicitely taken into account
when we will quantify the accuracy of our method using simulations.
Neglecting cross terms of mask and cosmological signal, the average
observed density is:
\be
\frac{\langle n_{\rm o}\rangle}{\Phi L_0} = S_A - \langle\epsilon\rangle 
- \left(S_B-\frac{1}{2}\,\right) \langle\epsilon^2\rangle +O(\epsilon^3)\, ,
\label{eq:nobsbar}
\e
so that the average density is affected by the foreground already at first
order in $\epsilon$ if the perturbation does not average to zero. The
second-order effect is due to the curvature of the luminosity function
at the luminosity threshold, that gives an imbalance between galaxies
that get in or out of the selection.

The resulting galaxy density contrast can be written as the sum of a
cosmological ($T_{\rm c}\delta$) and a non-cosmological ($T_{\rm nc}$) term:
\bea
  \delta_{\rm o} &=&  \frac{n_{\rm o}-\langle n_{\rm o}\rangle}{\langle n_{\rm o}\rangle}= 
  T_{\rm nc} + T_{\rm c} \, \delta\, ,\\
  T_{\rm nc} &=& - \frac{1}{S_A}(\epsilon-\langle\epsilon\rangle )
               -\frac{1}{S_A^2}\langle\epsilon\rangle(\epsilon-\langle\epsilon\rangle) 
               - \frac{S_B-\frac{1}{2}}{S_A}(\epsilon^2-\langle\epsilon^2\rangle) + O(\epsilon^3)\, ,\nonumber\\
  T_{\rm c} &=& b_1 - \frac{\beta_1\epsilon-b_1\langle\epsilon\rangle}{S_A}
  - \frac{\left(S_B-\frac{1}{2}\right)\left(\beta_1+\beta_1'\frac{\Phi}{\Phi'}\right)}{S_A}\,\epsilon^2
 \nonumber\\ & & 
  -\frac{\beta_1}{S_A^2}\,\langle\epsilon\rangle\,\epsilon
  + \frac{1+S_A\left(S_B-\frac{1}{2}\right)}{S_A^2}b_1\langle\epsilon^2\rangle +O(\epsilon^3)\, .\nonumber
  \label{eq:deltaobs}
\eea If no foreground is present then $\epsilon=0$ and the
non-cosmological term $T_{\rm nc}$ vanishes as expected. We will
consider later a (toy-model) case where bias is independent of
luminosity; in this case $b_1=\beta_1$ and $\beta_1'=0$, and
equation~\ref{eq:deltaobs} can be simplified to take the form
compatible with equation 2.19 of C17 and with the form assumed to
obtain equation~\ref{eq:start}: \be 1+\delta_{\rm o} = (1-A)
(1+\delta_{\rm g})\, ,
\label{eq:nolumdep}
\e
where
\be
  A =  \frac{1}{S_A}(\epsilon-\langle\epsilon\rangle)
         +\frac{1}{S_A^2}\langle\epsilon\rangle(\epsilon-\langle\epsilon\rangle) 
         +\frac{S_B-\frac{1}{2}}{S_A}(\epsilon^2-\langle\epsilon^2\rangle) + O(\epsilon^3)\, .
\e
In this case $T_{\rm nc}=-A$ and $T_{\rm c}=1-A$.

\subsection{Milky Way extinction}

As a prototype of foreground contamination that modulates survey
depth $\delta L(\boldsymbol\theta)$, we choose Milky Way extinction.
Extinction at a wavelength $\lambda$, quantified as a magnitude, is
denoted as $A_\lambda=R(\lambda) E(B-V)$, where $E(B-V)$ is the
reddening and $R(\lambda)$ an extinction curve. Reddening only depends
on the sky position, so we define the ``visibility mask'' ${\cal
  M}(\boldsymbol\theta):=E(B-V)$ as the reddening. We will assume (see
next section) that our Euclid-like catalogs consist of galaxies that
are detected in $H\alpha$ emission, so the impact of extinction on
galaxy selection acquires a redshift dependence.
The luminosity limit of the survey is:
\be
L_{\rm lim}(\boldsymbol\theta,z) = L_0 10^{0.4 R(z) {\mathcal M}(\boldsymbol\theta)} \, ,
\label{eq:extinction}
\e
where $R(\lambda)$ is computed at the wavelength of the $H\alpha$ line
at redshift $z$, $6563\times(1+z)$~{\AA}. It is useful to define the
scaling function:
\be
  S_C(z)=0.4 \ln 10\, R(z)\, .
\label{eq:S_Cz}
\e
Extragalactic surveys usually observe sky areas with $E(B-V)\la0.1$,
so, for a typical extinction curve, the largest values of $S_C{\cal M}$
will range from $\sim0.2$ in the $R$ band ($z\sim0$) to $\sim0.03$ in
the $K$ band ($z\sim2.4$). The relative variation of flux limit
$\epsilon$ is then worked out to second order as:
\be
\epsilon(\boldsymbol\theta,z) = {\rm e}^{ S_C(z)\, {\cal M}(\boldsymbol\theta)} -1
= S_C(z){\cal M}(\boldsymbol\theta) + \frac{1}{2} S_C^2(z){\cal M}^2(\boldsymbol\theta) + O({\cal M}^3)\, ,
\label{eq:epsilonM}
\e

The observed density contrast is thus expressed as:
\bea
  \delta_{\rm o} &=& T_{\rm nc} + T_{\rm c} \, \delta\, ,   \label{eq:deltaobs_m}\\
  T_{\rm nc} &=& - \frac{S_C}{S_A}({\cal M}-\langle{\cal M}\rangle)
    -\frac{S_C^2}{S_A^2}\langle{\cal M}\rangle({\cal M}-\langle{\cal M}\rangle)
    -\frac{S_BS_C^2}{S_A}({\cal M}^2-\langle{\cal M}^2\rangle)+ O({\cal M}^3)\, , \nonumber\\
  T_{\rm c} &=& 
  b_1 - \frac{S_C}{S_A}(\beta_1{\cal M}-b_1\langle{\cal M}\rangle)
    - \frac{S_BS_C^2}{S_A}(\beta_1{\cal M}^2-b_1\langle{\cal M}^2\rangle)
    -\frac{S_C^2}{S_A^2}\beta_1{\cal M}\langle{\cal M}\rangle \nonumber \\
    && - \frac{\left(S_B-\frac{1}{2}\right)S_C^2}{S_A} \beta_1' \frac{\Phi}{\Phi'} {\cal M}^2
    +\frac{S_C^2}{S_A^2}b_1\langle{\cal M}^2\rangle+ O({\cal M}^3)\, . \nonumber
\eea
For a bias that is independent of luminosity:
\be
  1+\delta_{\rm o}  =  (1-A) (1+\delta_{\rm g})\, , \label{eq:nolumdep_m}\,
  \e
  with
\be
  A = \frac{S_C}{S_A}({\cal M}-\langle{\cal M}\rangle)
    +\frac{S_C^2}{S_A^2}\langle{\cal M}\rangle({\cal M}-\langle{\cal M}\rangle)
    +\frac{S_BS_C^2}{S_A}({\cal M}^2-\langle{\cal M}^2\rangle)+ O({\cal M}^3)\, .
\e
This clarifies the reason why we added $1/2$ to the definition of
$S_B$ in equation~\ref{eq:SBz}: it was done to get here a more compact
form for $T_{\rm nc}$. It is worth noticing here that the three
functions $S_A$, $S_B$ and $S_C$ for a survey can be computed from the
galaxy luminosity function and from the extinction curve, and so it is
possible to assume them as known functions. We will discuss the role
of the extinction curve in Section~\ref{sec:testing}.

It is important to stress the domain of validity of
equation~\ref{eq:deltaobs_m}: we have assumed no correlation between
$\epsilon(\boldsymbol\theta)$ and $\delta({\mathbf x})$ and we have
neglected residual correlations due to the fact that volume averages
are not ensemble averages; we have assumed no environmental dependence
of the shape of the luminosity function, a linear bias scheme and a
second-order expansion of the integral and of the exponential that
relates $\epsilon$ and ${\cal M}$. Next step, carried out in
Section~\ref{sec:estimators}, is to construct estimators that can
average out the cosmological term $T_{\rm c}\delta$ and single out the
mask-dependent term $T_{\rm nc}$ in that equation.

Real surveys will be affected by several foregrounds at the same time,
so what one would measure is their combined effect. However, the way
different foregrounds impact as a function of redshift, i.e. their
$S_C(z)$ functions, may be different. For instance, a photometric
selection based on a single apparent magnitude, or a
wavelength-independent modulation of the photometric zero-point, would
give a constant $S_C$ function. The present method should then be
generalized to a sum of contributions, grouped according to the shape
of their $S_C(z)$ functions. For sake of simplicity we will not
consider this extension in this paper.

\section{Simulated catalogs}
\label{sec:mocks}

To build and test our method, we generated a set of 20 large simulated
catalogs of dark matter halos on the observer past light cone. The
catalogs were generated with the \pin\footnote{\tt
  http://adlibitum.oats.inaf.it/monaco/pinocchio.html} approximate
method \cite{Monaco2013,Munari2017}; see \cite{Monaco2016} for a
review on this technique.

{\pin} is a parallel code that implements the following semi-analytic
scheme: starting from a linear density contrast field sampled on a
regular cubic grid (as in the construction of initial conditions for
an N-body simulation), it predicts, for each grid element (or
particle), the time at which it is expected to collapse (to suffer
orbit crossing). This is done by smoothing the linear density contrast
on a set of scales, computing the collapse time using ellipsoidal
collapse as in \cite{Monaco1997}, and then taking for each particle
the earliest obtained collapse time. Collapsed particles are then
grouped into halos with an algorithm that mimics their hierarchical
assembly. Halo positions are computing using 3rd-order Lagrangian
Perturbation Theory \cite{buchert1992,bouchet1992,catelan1995}.

With the latest version of the code, we run 20 independent
realizations of a box of 3.2 Gpc$/h$, sampled with $4096^3$ particles.
We used a WMAP cosmology with $\Omega_0=0.285$,
$\Omega_\Lambda=0.715$, $\Omega_b=0.044$, $h=0.695$, $\sigma_8=0.828$,
$n_s=0.9632$. Each particle has a mass of $3.77\times10^{10}\ {\rm
  M}_\odot/h$ and, as we identified halos of at least 20 particles,  the smallest
resolved halo has mass $7.54\times10^{11}\ {\rm M}_\odot/h$. Halos
were output on the past light cone, from $z=2.5$ to
$z=0$, on a comoving volume given by the
projection of a sky circular area with semi-aperture of $60^\circ$,
thus covering $1/4$ of the sky. The box was replicated using periodic boundary conditions to
fill the cone volume.

We produced a sample of mock galaxy catalogs that broadly mimic the
spectroscopic survey of the forthcoming Euclid satellite. This will
observe star-forming galaxies through slit-less spectroscopy over
disconnected regions covering $\sim1/3$ of the sky (separated by the
Milky Way and the ecliptic). Such galaxies will mainly be detected through
their $H\alpha$ emission line (blended with $[NII]$), in a redshift
range that has been revised (with respect to the initial red
book \cite{laureijs2011} assumption) to be from 0.9 to 1.75. We will make use of a
larger redshift range, in order to have a larger number of independent
redshift bins. Indeed, the amount of information available to
reconstruct the mask will not be limited to this spectroscopic sample,
it will be possible to apply the same technique to the photometric
sample, to more distant emission line galaxies (e.g. Lyman-$\alpha$
emitters) and to quasars, thus justyfing the first
tests of the method on a larger redshift range. This point will be
further discussed in the conclusions.

Assuming a one-to-one relation between halo mass and galaxy luminosity
in the $H\alpha$ line (as in a naive Halo Occupation Distribution
model - HOD - with one single galaxy in each halo), halos were
number-matched to galaxies using the luminosity function of model 1 of
Pozzetti et al. (2016) \cite{Pozzetti2016}. This was done by dividing
the redshift range from 0 to 2.5 in bins of width $\delta z=0.05$; for
each bin the cumulative mass function of halos was computed by
averaging over the 20 light cones. Since halo masses take discrete values
multiples of the particle mass, we
redefine  halo masses by assuming that a halo of $N_p$ particles of
mass $M_p$ has a mass distributed in the interval from $N_pM_p$ to
$(N_p+1)M_p$, with a mass function equal to a power-law interpolation
of the numerical halo mass function at the same mass and redshift bin.
The relation between halo mass $M$ and $H\alpha$ luminosity $L$ was
then found by tracking at which $M$ and $L$ values the cumulative mass
and luminosity functions give the same number density. The matching
luminosities were interpolated with polynomials in mass and redshift,
to construct a function that assigns a luminosity to each halo. This
interpolation induces very little differences in the luminosity
function at a redshift around $z=1.3$, where the $\Phi_\star$
parameter of the Schechter fit in model 1 of Pozzetti has a break.
These differences are far smaller than the $\sim0.2$ uncertainty in
the number density of $H\alpha$ galaxies.

With this naive HOD model we obtained 20 mock galaxy catalogs covering
$1/4$ of the sky, with known angular position, observed redshift
(including redshift-space distortions) and $H\alpha$ luminosity
matching model 1 of Pozzetti. We assumed that the redshift is known
without observational uncertainty. Mock galaxy samples were selected
using a flux limit of $2\times10^{-16}\ {\rm erg\ s}^{-1}\ {\rm
  cm}^{-2}$. As a matter of fact, the halo mass limit of 20 particles
sets a redshift-dependent luminosity completeness, that may be higher
than the luminosity corresponding to the flux limit; in this case the
sample is incomplete and this happens at $z<0.75$, so we will consider
in this paper the redshift range from $z=0.8$ to $z=2.4$. One obvious
problem of these mock catalogs is that, assuming a one-to-one relation
between halos and galaxies, the 1-halo term that dominates clustering
on small scales is completely missing. Because we are mostly focused
on very large scales, this issue is not a worry in this context.

Due to the luminosity dependence of (linear) bias, the $T_{\rm c}$
term in equation~\ref{eq:deltaobs_m} for the observed density contrast
is complicated by the presence of three bias terms $b_1$, $\beta_1$
and $\beta_1'$. These do not enter in the reconstruction of the
visibility mask, that is based on $T_{\rm nc}$ in the same equation,
so their relevance is limited. However they introduce a further level
of complication in the analysis, that is convenient to avoid at this
stage. We decided to remove the luminosity dependence of bias with the
following procedure: for each redshift bin used in the abundance
matching procedure, we shuffle the values of halo masses, and
$H\alpha$ luminosities, among all the halos. The halo mass function is
unchanged, and so is the abundance matching procedure. This way, a
selection in luminosity will imply a random sampling of the halos, so
the selected sample will have the same linear bias $b_1$ independent
of luminosity. As a consequence, equation~\ref{eq:nolumdep_m} will
apply in place of equation~\ref{eq:deltaobs_m}. We however warn the
reader that bias generally depends on luminosity, so these mock catalogs
can be considered only as a convenient toy model of galaxy clustering.

For the mask, we considered the three extinction maps described above,
SFD, P13 and P15. All maps are provided to the community as HEALPIX
maps of either $E(B-V)$ reddening (SFD and P13) or $A_V$ extinction.
SFD is provided in a pixelization with NSIDE=512 while Planck maps are characterised by the
higher resolution of NSIDE=2048. As large scales are of interest here,
we resampled Planck maps to NSIDE=512 and used $R_V=3.1$ to relate
reddening and extinction: $A_V= 3.1 E(B-V)$. We will use P13 as our
reference visibility mask:
${\cal M}_{\rm true} = {\cal M}_{\rm P13} = E(B-V)_{\rm P13}$. For the
extinction curve $R(\lambda)$ we adopted the determination of
\cite{Cardelli89}. To avoid the plane of the Milky Way the axis of the
light cone was oriented toward the north galactic pole. Mocks were
masked by imposing a different limiting flux in each healpix sky cell.
Galaxy catalogs are divided into $N_z=17$ redshift bins, centered
around $z_i = 0.8 + i \Delta z$, with $i$ ranging from 0 to $N_z-1$
and $\Delta z = 0.1$, twice the value used for abundance matching. The
total redshift range is thus limited to $0.75<z<2.45$. Maps of galaxy
density contrast $\delta_{\rm o}$ were computed, for each redshift bin, by
counting galaxies in each (equal area) sky pixel covered by our
survey, using the same NSIDE=512 healpix map used for the mask, then
dividing the counts by the mean number of galaxies per pixel and
subtracting $1$. This was done for both the unmasked and the masked
versions of the catalogs.

\begin{figure}
\centering{\includegraphics[width=0.9\textwidth]{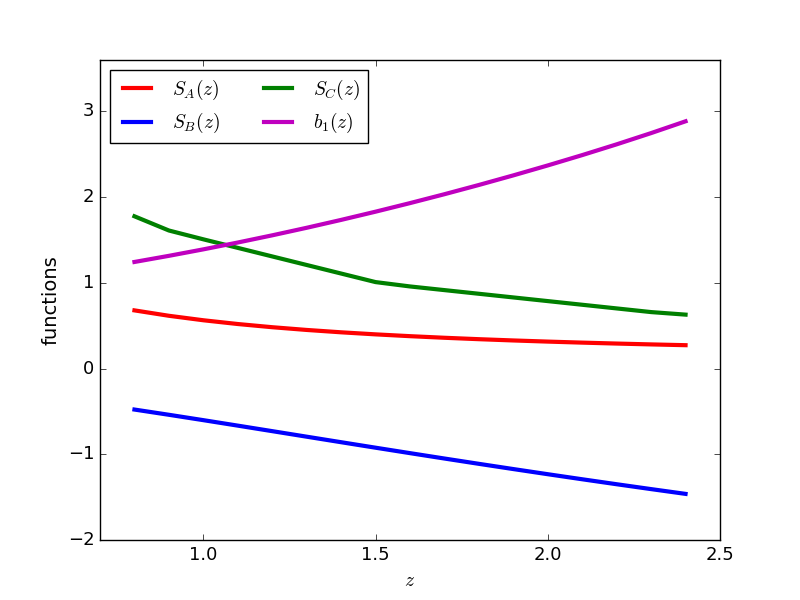}}
\caption{Scaling functions $S_A(z)$, $S_B(z)$ and $S_C(z)$ for our
  sample selection, and linear galaxy bias $b_1(z)$, computed for
  halos with mass $M\ge20M_p$.}
\label{fig:functions}
\end{figure}

Figure~\ref{fig:functions} shows the three scaling functions,
$S_A(z)$, $S_B(z)$ and $S_C(z)$, computed for our catalog selection,
based on the assumed luminosity function and extinction curve. The
figure shows also the value of the linear bias $b_1(z)$, measured at a
set of redshifts (where outputs for the whole periodic simulation box
were available) by comparing the halo power spectrum, measured with
the code of  \cite{sefusatti2016}, with the linear
power spectrum; values at other redshifts are obtained by
interpolation. This bias is computed for all the halos with $M\ge
20M_p$, so it applies to the catalogs obtained with shuffled halo
masses and is independent of luminosity.

\section{Reconstruction of the visibility mask}
\label{sec:reconstructions}

\subsection{Estimators of the mask-dependent term in the observed
  density contrast}
\label{sec:estimators}

The non-cosmological term $T_{\rm nc}$ of equation~\ref{eq:deltaobs_m}
depends only on the visibility mask, and can be singled out by
averaging, pixel by pixel, the observed density measurement
$\delta_{\rm o}$ over several redshift bins, under the assumption that
the cosmic density contrast $\delta$ averages to zero, at least in an
ensemble-average sense, along the same line of sight. This averaging
will be imperfect because it will be carried out over a limited volume
(and with redshift-dependent weights given by $T_{\rm c}$), and not
over a large number of realizations; in this paper we will carefully
quantify this residual by applying the method to our set of mock
catalogs.

In $T_{\rm nc}$ the mask appears through the terms $\langle{\cal M}\rangle$,
$({\cal M} - \langle{\cal M}\rangle)$ and $({\cal M}^2 - \langle{\cal
  M}^2\rangle)$. The inversion of this relation is affected
by multiple solutions at low mask values, the most interesting regions
for an extragalactic survey. However, prior knowledge of the first two
moments of the mask, $\langle{\cal M}\rangle$ and $\langle{\cal
  M}^2\rangle$, allows to recast the problem in a more convenient way.
In the following we will assume that $\langle{\cal M}\rangle$ and
$\langle{\cal M}^2\rangle$ are known, we will demonstrate in
Section~\ref{sec:moments} that these moments can be worked out to a
  sufficient accuracy, either by using external data or by internally
  estimating them from the galaxy survey.

With prior knowledge of the first two moments of the mask, it is
possible, using equation~\ref{eq:nobsbar}, to recover the true average
galaxy density $\langle n_{\rm g} \rangle$ from the observed $\langle n_{\rm
  o}\rangle$ and from the $S_B$ and $S_C$ scaling functions:

\begin{equation}
\frac{\langle n_{\rm g} \rangle}{\Phi L_0} = \frac{\langle n_{\rm o} \rangle}{\Phi L_0} + S_C(z) 
\langle {\cal M} \rangle + S_B(z) S_C(z)^2 \langle {\cal M}^2 \rangle\, .
\label{eq:n_from_no}\end{equation}

\noindent
It is useful to define a rescaled density contrast $\delta_{\rm r}$
as:
\begin{multline}
\delta_{\rm r} := \frac{n_{\rm o}}{\langle n_{\rm g}\rangle} -1 = T_{\rm nc} + T_{\rm c} \, \delta =
-\left[ \frac{S_C}{S_A}{\cal M}  + \frac{S_BS_C^2}{S_A}{\cal M}^2\right]
\\
+ \left[ b_1 - \frac{S_C}{S_A}\beta_1 {\cal M} -
  \frac{S_BS_C^2}{S_A}\beta_1 {\cal M}^2
+\frac{\left(S_B-\frac{1}{2}\,\right)S_C^2}{S_A}\beta_1'\frac{\Phi}{\Phi'}{\cal
  M}^2 \right]\times \delta + O({\cal M}^3)
\, .
\label{eq:deltaobs_resc}\end{multline}

\noindent
When bias is independent of luminosity:
\begin{equation}
1+\delta_{\rm r} = (1-A) (1+\delta_{\rm g})\, , \ \ \ \ \ \ \ \ \ \ 
A  = \frac{S_C}{S_A}{\cal M}  + \frac{S_BS_C^2}{S_A}{\cal M}^2\, .
\label{eq:deltar2}\end{equation}

In what follows we denote the values of the $S_A$, $S_B$ and $S_C$
functions, evaluated at the central redshift $z_i$ of the $i-$th bin,
as $S_{Ai}=S_A(z_i)$, $S_{Bi}=S_B(z_i)$ and $S_{Ci}=S_C(z_i)$. We
define two estimators that average out, in an ensemble average sense,
the cosmological signal and single out the redshift average of $T_{\rm
  nc}$ term, that depends only on the mask ${\cal M}$. The average
density estimator $E_{\rm av}(\boldsymbol\theta)$ in each sky pixel is
defined as:

\begin{equation}
E_{\rm av}(\boldsymbol\theta) \equiv 
\frac{1}{N_z} \sum_i \delta_{\rm r}([z_i,\boldsymbol\theta])
\simeq 
- \frac{1}{N_z} \sum_i \frac{S_{Ci}}{S_{Ai}}
 {\cal M}
- \frac{1}{N_z} \sum_i \frac{S_{Bi}S_{Ci}^2}{S_{Ai}}
 {\cal M}^2 + O({\cal M}^3)\, .
\label{eq:Eav}\end{equation}

\noindent
The sum is over all $N_z$ redshift bins; here we have made it explicit
that the position $\mathbf{x}$ is decomposed into angular and redshift
coordinates: $\mathbf{x}=[z_i,\boldsymbol\theta]$. The square average
density estimator $E_{\rm sq}(\boldsymbol\theta)$ is defined as:

\begin{multline}
E_{\rm sq}(\boldsymbol\theta) \equiv 
\frac{1}{N_p} \sum_i \sum_{j>i}
\delta_{\rm r}([z_i,\boldsymbol\theta])\delta_{\rm r}([z_j,\boldsymbol\theta]) \\
\simeq  
\frac{1}{N_p} \sum_i \sum_{j>i} \frac{S_{Ci}S_{Cj}}{S_{Ai}S_{Aj}}   {\cal M}^2 +
\frac{1}{N_p} \sum_i \sum_{j>i} \frac{S_{Ci}S_{Cj}(S_{Bi}S_{Ci}+S_{Bj}S_{Cj})}{S_{Ai}S_{Aj}}   {\cal M}^3 
+O({\cal M}^4)
\, .
\label{eq:Esq}\end{multline}

\noindent
Here $N_p= N_z (N_z -1)/2$ is the number of independent pairs of
redshift bins, considering each pair only once; for $N_z=17$ we have
$N_p=136$. The sum in equation~\ref{eq:Esq} could be limited by
removing nearby redshift bin pairs, that may be affected by residual
cosmological correlation. We will show in Section~\ref{sec:cross} that
this correlation is marginal, in the following we keep all redshift
bin pairs to maximize the signal.

The quadratic and cubic algebraic equations~\ref{eq:Eav} and
\ref{eq:Esq} can be inverted to provide, for each pixel, estimates of
the mask term that we will call ${\cal M}_{\rm av}$ and ${\cal M}_{\rm
  sq}$. It is important to notice that the two mask reconstructions
are tightly correlated, we will later quantify and use this property.
The computation of the coefficients in equations~\ref{eq:Eav} and
\ref{eq:Esq} is independent of any clustering measurement, and depends
only on the knowledge of the luminosity function and extinction law.
The resulting quantities provide noisy reconstructions of the mask,
the noise being due to incomplete averaging-out of the cosmological
signal (including its shot noise). As a first step, we illustrate the
degree of recovery of the mask by applying it to a resampling of the
galaxy density field. The maps of density contrast $\delta_{\rm
  o}(z_i)$, computed as described in Section~\ref{sec:mocks}, are
resampled to a healpix NSIDE=64 grid, and rescaled to obtain
$\delta_{\rm r}(z_i)$ (equation~\ref{eq:deltaobs_resc}) using for
$\langle{\cal M}\rangle$ and $\langle{\cal M}^2\rangle$ the values
computed using the true P13 mask at the same resampling. Densities
$\delta_{\rm r}(z_i)$ are then used to compute the two estimators
$E_{\rm av}$ and $E_{\rm sq}$. Finally, equations~\ref{eq:Eav} and
\ref{eq:Esq} are numerically inverted to obtain, pixel by pixel, the
mask reconstructions ${\cal M}_{\rm av}$ and ${\cal M}_{\rm sq}$.

\begin{figure}
\centering{\includegraphics[width=0.49\textwidth]{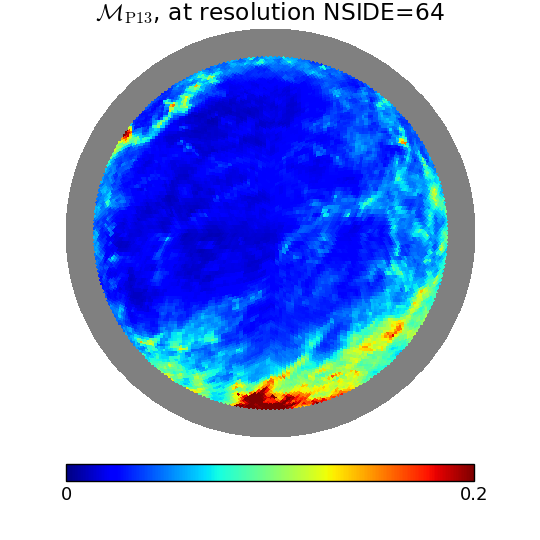}} \\
\centering{
\includegraphics[width=0.49\textwidth]{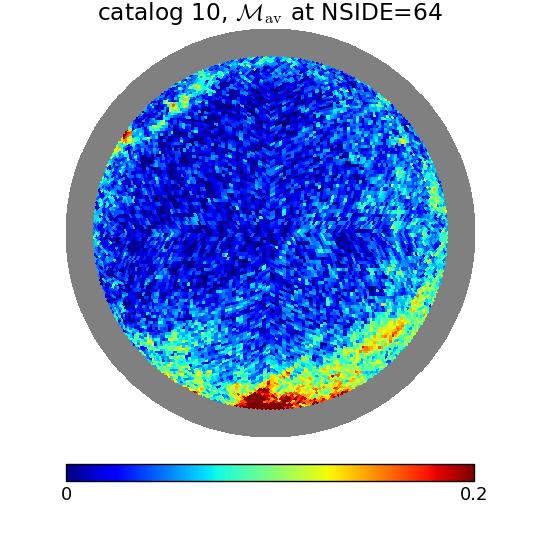}
\includegraphics[width=0.49\textwidth]{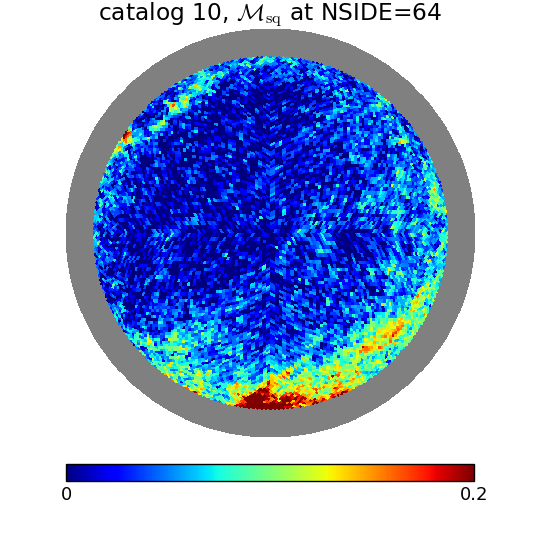}}
\caption{P13 reddening ${\cal M}_{\rm P13} = E(B-V)_{\rm P13}$ used to mask the mock
  catalogs (upper panel), and reconstructions ${\cal M}_{\rm av}$ (middle panel) and
  ${\cal M}_{\rm sq}$ (lower panel) based on $E_{\rm av}$ and $E_{\rm sq}$ estimators. All
  maps are orthogonal projections of the survey, a cone of $60^\circ$
  of semi-aperture located at the north galactic pole, and are shown
  at a healpix resolution of NSIDE=64.}
\label{fig:maps}
\end{figure}

Figure~\ref{fig:maps} shows, for one mock catalog, orthogonal
projections of maps of the true (P13) visibility mask and of the
reconstructions ${\cal M}_{\rm av}$ and ${\cal M}_{\rm sq}$. The main large-scale features
are neatly recovered, though noise is evident. A more quantitative
analysis reveals that both reconstructions give a nearly unbiased estimate of
the true mask for ${\cal M} \le0.1$, with a scatter that at this
resolution amounts to $\sim$0.03 mag.

It is worth stressing that this level of agreement is obtained thanks
to the second-order expansion of the integral in
equation~\ref{eq:expanded}, the first-order expansion gives a much
more biased estimate of the mask. A third-order expansion would likely
improve the results at ${\cal M}>0.1$, but real large-scale surveys
tend to avoid such relatively high extinction regions.

\subsection{Combining ${\cal M}_{\rm av}$ and ${\cal M}_{\rm sq}$ into a best reconstruction of the mask}
\label{sec:bestreconstruction}

\begin{figure}
\begin{center}
\includegraphics[width=0.9\textwidth]{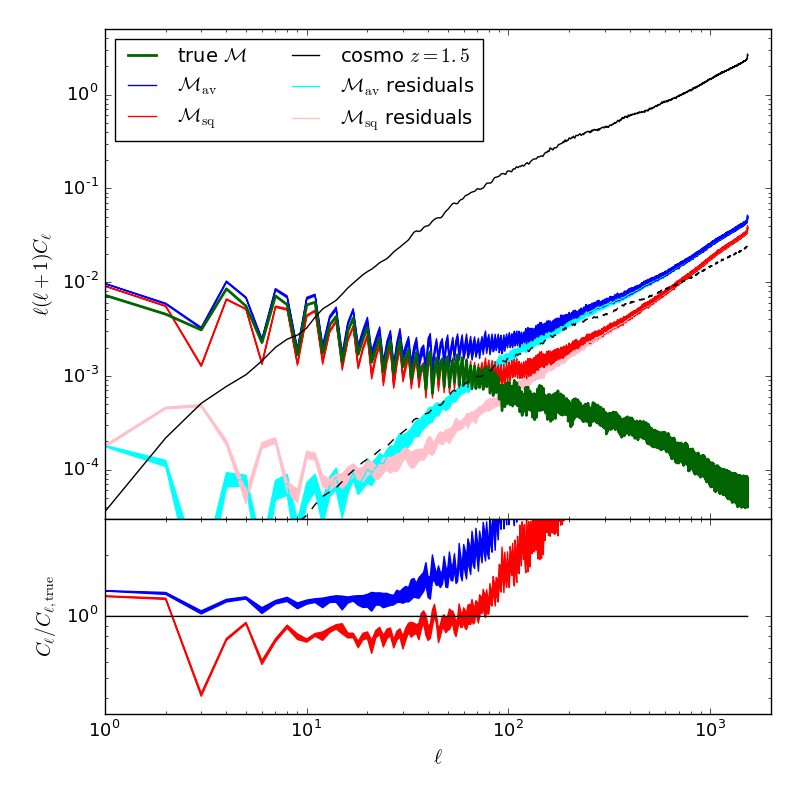}
\end{center}
\caption{Angular power spectra of the true mask ${\cal M}_{\rm true}$
  (green curve) and its reconstructions ${\cal M}_{\rm av}$ and ${\cal
    M}_{\rm sq}$ (blue and red curves). Reconstructions are averaged
  over the 20 light cones, the line width marks the standard deviation
  of the obtained clustering measurements. The cyan and pink lines
  give the power spectrum of the reconstruction residuals ${\cal
    M}_{\rm av} - {\cal M}_{\rm true}$ and ${\cal M}_{\rm sq} - {\cal
    M}_{\rm true}$. The black continuous line gives, as a reference,
  the measured clustering signal of galaxies at $z=1.5$, averaged over
  the 20 cosmological, unmasked mocks, and including shot noise. The
  dashed black line is $1/100$th of the previous curve. The lower
  panel gives the ratio between the two reconstructions and the true
  mask, and between the P15 and SFD masks and the true (P13) mask.}
\label{fig:clmask}
\end{figure}

When applied to the density field computed on the native NSIDE=512
healpix grid, the two reconstructions show a much higher level of
scatter at the pixel-by-pixel level. It is very instructive to analyse
their angular power spectra. To this aim, we used the 20 mock light
cones constructed with the procedure described in
Section~\ref{sec:mocks}. Using the procedure described above in
Section~\ref{sec:mocks}, we computed the angular density maps in
redshift bins of width $\Delta z=0.1$ on a healpix grid of NSIDE=512,
computed from these the $E_{\rm av}$ and $E_{\rm sq}$ estimators and
used them to obtain ${\cal M}_{\rm av}$ and ${\cal M}_{\rm sq}$ by
inverting equations~\ref{eq:Eav} and \ref{eq:Esq}. We used the
implementation of the ANAFAST code provided by the Healpy
package to compute the $C_\ell$ angular power spectra of the three
reddening mask presented above and, for each mock, of the two
reconstructed masks ${\cal M}_{\rm av}$ and ${\cal M}_{\rm sq}$. To
account for the simple survey geometry, that encompasses $1/4$th of
the sky, we multiplied all results by a factor of $4$.
Figure~\ref{fig:clmask} shows the results in terms of $\ell (\ell+1)
C_\ell$. The green, blue and red lines give respectively the angular
power spectra of the P13 ``true'' mask ${\cal M}_{\rm true} =
E(B-V)_{\rm P13}$ and its two ${\cal M}_{\rm av}$ and ${\cal M}_{\rm
  sq}$ reconstructions; these are denoted by a band as thick as the
standard deviation of measurements over the 20 realizations. The bands
with lighter colors give the power spectra of reconstruction
residuals, ${\cal M}_{\rm av}-{\cal M}_{\rm true}$ and ${\cal M}_{\rm
  sq}-{\cal M}_{\rm true}$. As a reference, we report in this plot (as
a black continuous line) the measurement of angular clustering,
averaged over our 20 light cones, of galaxies in the cosmological,
unmasked samples at a fiducial redshift of $z=1.5$ (this signal
includes shot noise). Because of a fortunate combination of matter
clustering and halo bias, the level of this signal is pretty stable
with redshift, so this line is representative of the
expected level of clustering in the whole redshift range considered.
The black dashed line gives for reference $1/100$th of the
cosmological signal.

Both reconstructions reproduce the power spectrum of the mask at $\ell\la30$,
with a constant bias on large scales that amounts to $\sim15$\% for
${\cal M}_{\rm av}$ and $\sim25$\% for ${\cal M}_{\rm sq}$. At higher $\ell$'s the power
spectra of reconstructions nose up to roughly follow the shape of the cosmological
signal. The power spectra of reconstruction residuals are very informative to
analyse. On very large scales they run parallel to the mask spectrum,
while their raise at higher $\ell$-values is broadly parallel to the
cosmological signal. This can easily be interpreted as the effect of
incomplete averaging out of the cosmological signal. The $E_{\rm sq}$-based
reconstruction gives lower residuals because the averaging is performed not
over all redshift bins ($N_z=17$) but over all bin pairs ($N_p=136$)
thus reducing the residual cosmological signal by a factor of 3, to a
level that is more than a factor of $100$ below the cosmological
signal. The $E_{\rm av}$-based reconstruction gives percent-level residuals
for $\ell<100$.

\begin{figure}
\centering{
\includegraphics[width=0.49\textwidth]{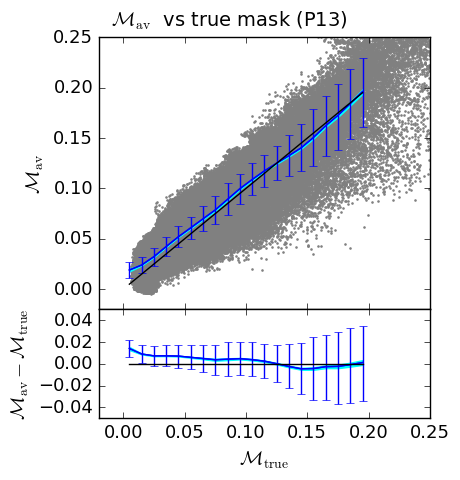}
\includegraphics[width=0.49\textwidth]{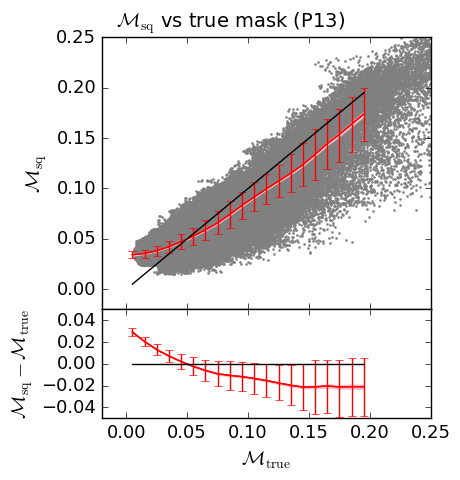}}
\centering{
\includegraphics[width=0.49\textwidth]{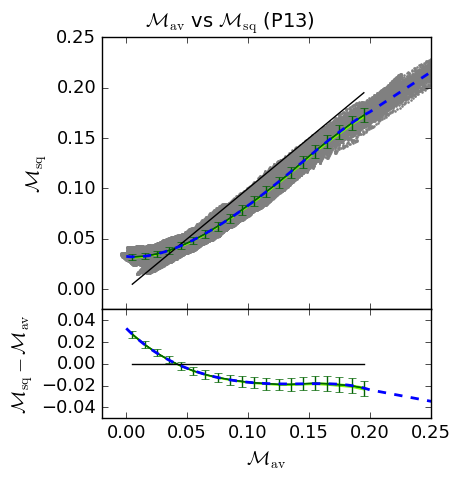}
\includegraphics[width=0.49\textwidth]{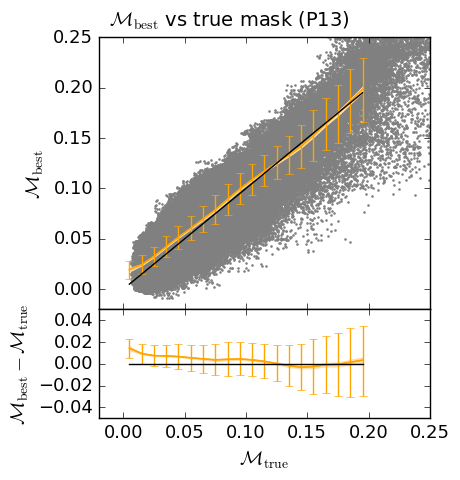}}
\caption{Pixel-by-pixel scatterplots of reconstructed and true masks. In all
  panels grey dots denote pixel values, lines the average in bins of
  the quantity on the x-axis, the lighter-colored area the sample
  variance on the mean, errorbars the 16th and 84th percentiles in the
  same bins, averaged over the 20 mocks. The lower sub-panels give the
  residuals over the bisector line, denoted as a black line; here
  points are not reported. Upper panels: correlation of smoothed
  ${\cal M}_{\rm av}$ and ${\cal M}_{\rm sq}$ versus the true mask. Lower panels:
  correlation of the two smoothed reconstructions (left) and correlation of the
  best reconstruction with the true mask (right).}
\label{fig:bestmodel}
\end{figure}

This result allows us to quantify the angular scales beyond which the
reconstruction power spectra are dominated by the cosmological signal
contamination. Notably, these scales will be valid as long as the
foreground that contaminates a given survey has a power spectrum
similar to the one used to create the mocks, that is biased toward
larger scales. Because this excess small-scale power is correlated
with the cosmological signal, it is important to filter it out. We do
this by multiplying the $a_\ell^m$ coefficients of the spherical
harmonics expansion of the reconstructed mask by a Gaussian smoothing
function of $\ell$:

\begin{equation}
(a_\ell^m)_{\rm smoothed,i} = (a_\ell^m)_i
  \exp\left(-\left(\frac{\ell}{\ell_{i}}\right)^2\right)\, .
\label{eq:filter}
\end{equation}

\noindent
where $i={\rm av}$ or sq. Here the smoothing angular scales for the ${\cal
  M}_{\rm av}$ and ${\cal M}_{\rm sq}$ reconstructions, $\ell_{\rm av}$ and $\ell_{\rm sq}$, are kept
separated to allow for a stronger filtering of the ${\cal M}_{\rm av}$ reconstruction,
that is more contaminated by the cosmological residual. Good values
for these smoothing scales are found to be $\ell_{\rm av}=90$ and
$\ell_{\rm sq}=120$. In the following, we will only use the smoothed versions
of the $E_{\rm av}$- and $E_{\rm sq}$-based reconstructions.

\begin{figure}
\centering{
\includegraphics[width=0.49\textwidth]{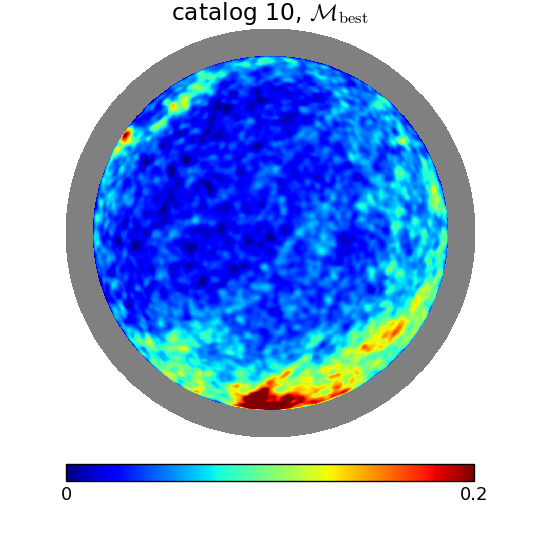}
\includegraphics[width=0.49\textwidth]{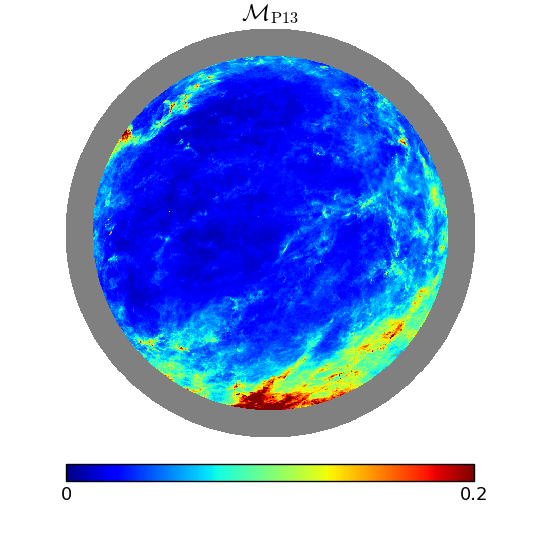}
}
\caption{Orthogonal projections of the best reconstruction ${\cal M}_{\rm
    best}$ for the reconstructed mask (left panel), obtained with one
  of the mock catalogs, and of the true P13 reddening map used to mask
  the mock galaxy catalogs (right panel).}
\label{fig:bestmap}
\end{figure}

\begin{figure}
\begin{center}
\includegraphics[width=0.9\textwidth]{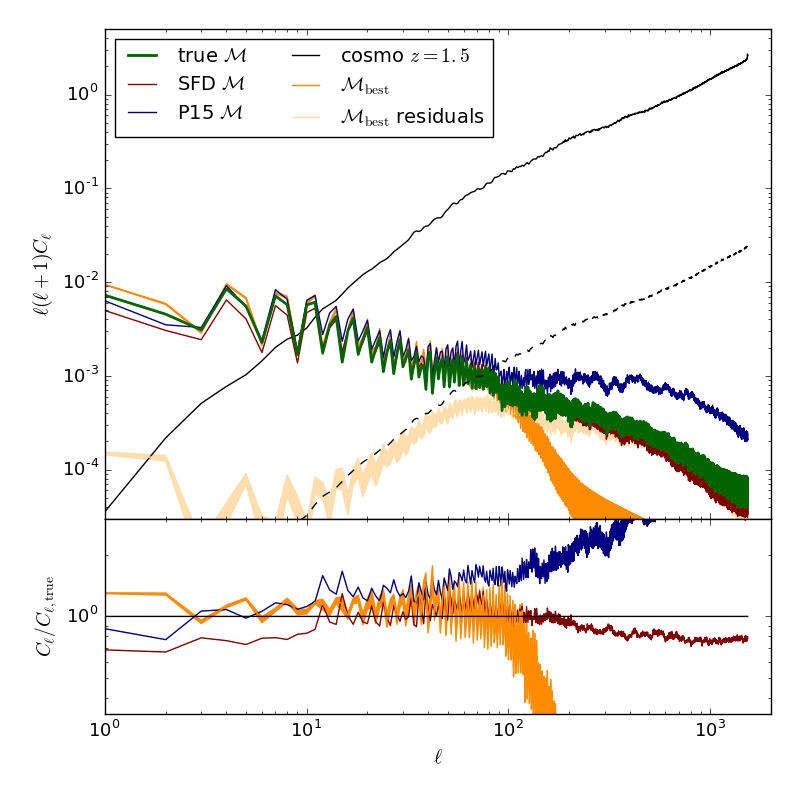}
\end{center}
\caption{Angular power spectra of the true mask ${\cal M}_{\rm true}$
  (green curve) and of the best reconstruction ${\cal M}_{\rm best}$ (orange
  curve). Reconstructions are averaged over the 20 light cones, the line width
  marks the standard deviation of the obtained clustering
  measurements. The light orange line gives the power spectrum of the
  reconstruction residuals ${\cal M}_{\rm best} - {\cal M}_{\rm true}$. The
  black continuous line gives, as a reference, the measured clustering
  signal of galaxies at $z=1.5$, averaged over the 20 cosmological,
  unmasked mocks. The dashed black line is $1/100$th of the previous
  curve. The lower panel gives the ratio between the reconstructions and the
  true mask.}
\label{fig:clbest}
\end{figure}

Figure~\ref{fig:bestmodel} shows, in the two upper panels, the
correlation between the two smoothed versions of the reconstructions
with the true visibility mask ${\cal M}_{\rm true}$. Here the grey
points represent the values obtained in all pixels in the survey for
one single mock catalog; the lines give the mean of the reconstructed
mask in bins of ${\cal M}_{\rm true}$, averaged over the 20 mocks,
while the bands with lighter color around the mean give the sample
variance (standard deviation) of the mean. Errorbars give the scatter
of pixels in the same bins (quantified by the 16th and 84th
percentiles), averaged over the 20 mocks. Residuals over the bisector
line (denoted by a black line) are reported in the lower sub-panels.
${\cal M}_{\rm av}$ gives a remarkably unbiased reconstruction of the
mask, with a bias below 0.01 mag with the exception of the very first
point, and a scatter raising from 0.01 mag at low extinctions to 0.03
mag at ${\cal M}_{\rm true}=0.15$. ${\cal M}_{\rm sq}$ gives a more
biased reconstruction, with an overestimate of 0.03 mag at the lowest
reddening, turning negative beyond ${\cal M}_{\rm true}>0.05$; the
scatter is lower than the ${\cal M}_{\rm av}$ case. The flattening of
${\cal M}_{\rm sq}$ at low extinctions is mostly due to the quadratic
nature of the relation between the estimator $E_{\rm sq}$ and ${\cal
  M}_{\rm sq}$, that forces mask values to be positive.

It is convenient to combine the two reconstructions to obtain a nearly
unbiased estimate of the mask with low residuals. One way could be to
fix the bias of ${\cal M}_{\rm sq}$ by using the relation found with
mocks; the accuracy of this reconstruction would depend much on how
mocks have been constructed. A more conservative procedure can be
built on the basis that we can directly measure the relation between
the two reconstructions, and that this has a low scatter. This
relation is shown in the lower left panel of
figure~\ref{fig:bestmodel}. Using this relation, one can force one
reconstruction to have the same bias as the other one, and because
${\cal M}_{\rm av}$ is nearly unbiased, we use this relation to
transform ${\cal M}_{\rm sq}$. We fit the relation ${\cal M}_{\rm sq}
= f({\cal M}_{\rm av})$ with a polynomial (extrapolated as a linear
relation for ${\cal M}>0.2$), reported on the figure as a dashed line.
This fit is performed only once, we checked that $f({\cal M}_{\rm
  av})$ is the same for all cases considered in this paper. We then
correct ${\cal M}_{\rm sq}$ to have the same bias as ${\cal M}_{\rm
  av}$, that is known to be very low, by constructing the following
``best reconstruction'':

\begin{equation}
  {\cal M}_{\rm best} = {\cal M}_{\rm sq} - [f({\cal M}_{\rm av}) - {\cal M}_{\rm av}]
  \label{eq:bestmodel}
\end{equation}

\noindent
Figure~\ref{fig:bestmodel} gives, in the lower right corner, the
correlation of ${\cal M}_{\rm best}$ with the true mask. It shows how
this best reconstruction is biased in the same way as $M_{\rm av}$,
i.e. by less than 0.01 mag, with the only exception of the first bin.
The quality of the reconstruction can be appreciated in
Figure~\ref{fig:bestmap}, that shows the best reconstructed reddening
map obtained with one specific catalog beside the original P13 map at
the NSIDE=512 resolution used to mask the catalogs.

Figure~\ref{fig:clbest} shows the angular power spectrum of the best
reconstruction and its residual (the orange and lighter colored
bands), compared with that of the true mask (in green). We compare
these also to the angular power spectra of P15 and SFD masks (in dark
blue and maroon respectively), as the differences between the masks
can be taken as an order-of-magnitude indication of its uncertainty.
The angular clustering is recovered by the best model in an almost
unbiased way for $\ell<50$, while some gradual loss of power is
present up to $\ell\sim100$, where the reconstruction drops. This loss
is mostly due to the lower value of the $\ell_{\rm av}$ smoothing
scale; however, due to the higher contamination of ${\cal M}_{\rm
  av}$, a less strong smoothing would increase both power and
residuals. This choice keeps the power spectrum of residuals well
below 1\%, thus avoiding significant correlation of signal and noise.
If a better match of the power spectrum is needed, it would be better
to use these simulations to increase high-$\ell$ power, boosting both
signal and contamination, than increase $\ell_{\rm av}$, that would
boost only the contamination.

\section{Angular power spectra and cross correlations of density
  fields}
\label{sec:cl}

\subsection{The contribution of lensing}
\label{sec:lensing}

\begin{figure}
\centering{
\includegraphics[width=0.45\textwidth]{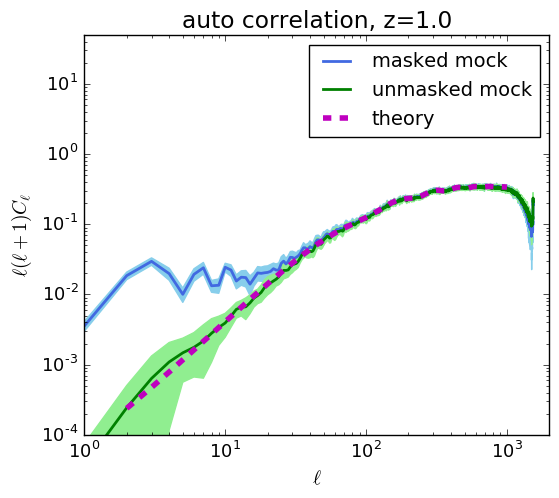}
\includegraphics[width=0.45\textwidth]{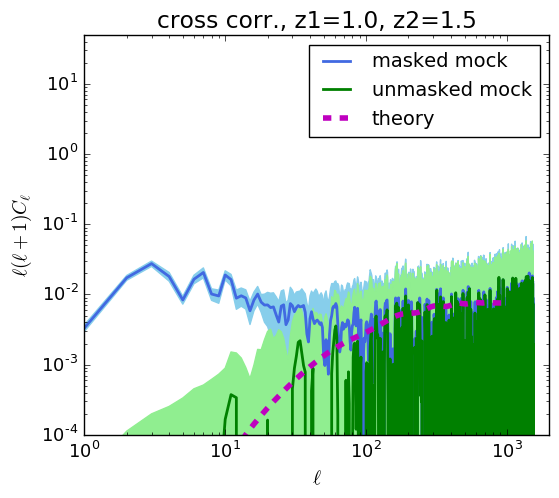}}
\centering{
\includegraphics[width=0.45\textwidth]{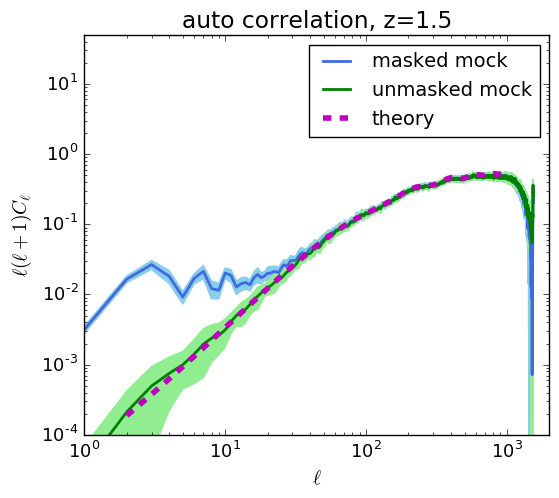}
\includegraphics[width=0.45\textwidth]{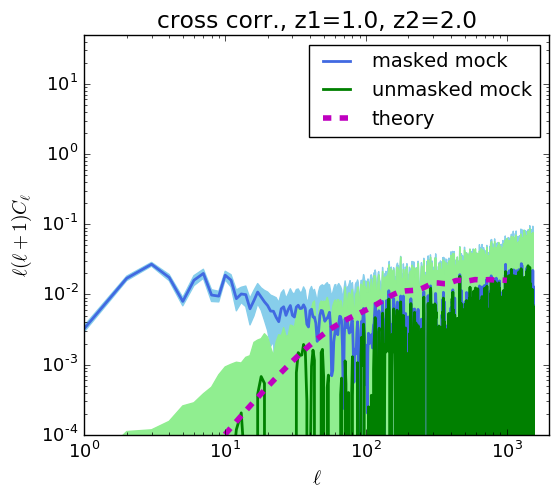}}
\centering{
\includegraphics[width=0.45\textwidth]{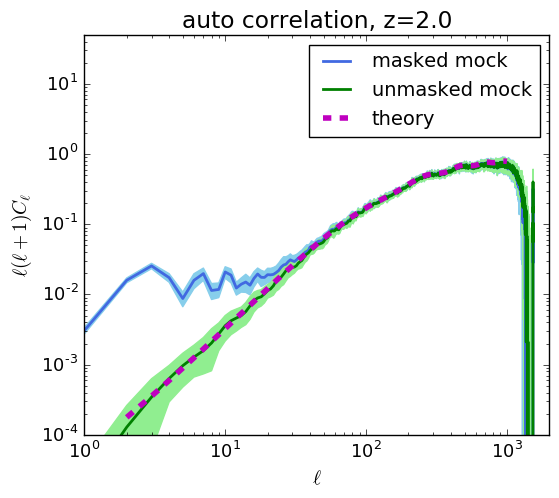}
\includegraphics[width=0.45\textwidth]{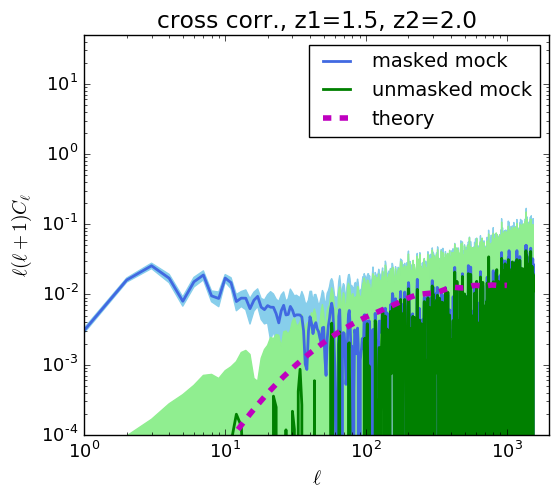}}
\caption{Left panels: angular correlation of the galaxy density field
  in redshift bins centered at $z=1$, $1.5$ and $2$, of width $\Delta
  z=0.1$. The green lines give the cosmological signal averaged over
  the 20 unmasked mock catalogs, the blue lines give the same
  measurement for catalogs masked with the P13 reddening map.
  Lighter-colored areas denote sample variance. The magenta dashed
  lines give the theoretical expectation including clustering,
  redshift-space distortions and lensing. Right panels: cross
  correlation of density fields at the same redshifts as above, with
  the same color coding.}
\label{fig:correlations}
\end{figure}

The reconstruction described in the previous sections relies on the
assumption that the cross-correlation of cosmological signal in
different redshift bins is vanishing. However, correlations in real
data will be induced by gravitational lensing, mostly through
magnification bias. This effect is not present in the mock catalogs
that we are using for this analysis. To understand the expected
influence of lensing, we show in figure~\ref{fig:correlations}
measured angular auto- (left panels) and cross- (right panels)
correlation functions of galaxies at redshifts $z=1.0$, $1.5$ and
$2.0$. In all panels the green and blue lines give the clustering of
unmasked and masked catalogs, averaged over the 20 realizations, while
the areas in lighter colors give the corresponding sample variance
(quantified by the standard deviation over the 20 mocks). In the left
panels the contribution of a Poissonian shot noise is subtracted from
the signal, estimated as $4\pi f_{\rm sky}/ N$, where $f_{\rm
  sky}=0.25$ is the fraction of the sky covered by the survey and $N$
is the number of galaxies in the redshift bin. The left-side column
shows that the auto correlations of unmasked catalogs agree very well,
as expected, with linear theory predictions, while masked catalogs
show excess power at the same scale where the mask correlation is
significant. On the right-side colum, the unmasked catalogs show
cross-correlations consistent with noise around a vanishing value,
while cross-correlations remain significant at $\ell<100$.

These measurements are compared with linear theory predictions, reported
in the plots as magenta dashed lines. 
Linear theory predictions have been computed with {\sc
  class}gal~\cite{Blas:2011rf,DiDio:2013bqa,DiDio:2016ykq}. The
redshift-dependent angular power spectra between $i$-th and $j$-th
redshift bins are obtained as

\be
c_\ell^{(ij)} = 4 \pi \int \frac{dk}{k} \Delta^{(i)}_\ell \left( k  \right)  \Delta^{(j)}_\ell \left( k  \right) \mathcal{P}_R \left( k \right)  
\e

\noindent
where $\mathcal{P}_R$ is the dimensionless primordial curvature power
spectrum and $\Delta_\ell^{(i)} \left( k \right)$ is the full angular
transfer function in the $i$-th redshift bin. In this work we consider
the main contributions, namely density, redshift space distortion
(RSD) and lensing convergence defined as follows:

\bea
\Delta_\ell^{\delta (i)} \left( k \right)  &=& \int dr \ W_i\left( r \right) b_1 T_\delta\left( k ,r \right) j_\ell \left( k r \right) \, ,
\\
\Delta_\ell^{{\rm rsd} (i)} \left( k \right)  &=& \int dr \  W_i\left( r \right) \frac{k}{a H} T_v\left( k ,r \right) j''_\ell \left( k r \right) \, ,
\\
\Delta_\ell^{\kappa (i)} \left( k \right)  &=& \ell \left( \ell + 1 \right) \int dr \ W_i \left( r \right)  \int_0^r dr'  \frac{2- 5 s}{2}  T_{\Phi + \Psi} \left( k , r' \right)  j_\ell \left( k r' \right)\, ,
\eea

\noindent
where the Fourier transfer functions $T\left( k ,r \right)$ are
normalized to the primordial curvature perturbation and they refer,
respectively, to density, velocity and Bardeen potentials, $j_\ell$
are the spherical Bessel functions and $j''_\ell$ their second
derivative with respect to the argument. The redshift binning is
denoted by the function $W_i$, normalized to unity. The shape of $W_i$
is assumed to take into account redshift errors; here we are assuming
no error on spectroscopic redshift, so we used a shape very near to a
square window function, checking that results are insensitive to
further sharpening.

Linear predictions depend on the galaxy bias $b_1(z)$ and the
magnification bias $s(z)$. Here the bias is assumed to be independent
of luminosity, so these predictions are comparable with our mocks
where the luminosity dependence of bias has been removed by shuffling
halo masses (see Section~\ref{sec:mocks}). Linear bias $b_1$ was
obtained as explained in Section~\ref{sec:mocks} and is reported in 
figure~\ref{fig:functions}, magnification bias $s(z)$ is defined as
the logarithmic slope of the luminosity function at the threshold
luminosity, so $s(z)=2S_B(z)-1$.

The upper panels show that the clustering of cosmological unmasked
catalogs follows very closely linear theory, despite missing the
lensing term. This is no surprise, because the contribution of lensing
(whose level is appreciable in the right panels) is negligible for
auto correlations with respect to clustering and RSD for redshift bins
of the width $\Delta z =0.1$ (see e.g.~\cite{Didio2}). This is not the
case for cross correlations, where the expectation for clustering and
RSD is very low, and the signal is dominated by lensing, that is
missing in the mock catalogs. For the masked mocks, expected lensing
and measured cross correlation cross at $\ell\sim 50-100$, after which
lensing follows the envelope of the wide fluctuations. This
illustrates once again that the importance of a foreground like Milky
Way extinction is limited to the largest angular scales.

\subsection{Reconstructing the true angular clustering}
\label{sec:correcting}

The customary way to correct for a given foreground is to create a
random catalog that is subject to the same selection bias as the
measured one. This would be straigthforward to implement, and would
make it possible to measure clustering in redshift space and assess
how well the cosmological clustering signal is recovered. We leave
this project to future work, and test here only the recovery of 
angular clustering. Equation~\ref{eq:deltaobs_m}, that relates the
observed density contrast with the matter one, can be inverted if the
mask ${\cal M}$ is known or a reconstruction for it is available. The
process is more complicated for the general case of
luminosity-dependent bias, but if luminosity dependence is absent, as
in our mock catalogs based on shuffled halo masses, then $b_1 =
\beta_1$ and $\beta_1'=0$, so the inversion is easily performed:

\begin{equation}
1+\delta_{\rm g} = \left[ 1 
  + \frac{S_C}{S_A}\left(1+\frac{S_C}{S_A}{\cal M}\right)({\cal M} - \langle{\cal M}\rangle)
  + \frac{S_BS_C^2}{S_A}({\cal M}^2 - \langle{\cal M}^2\rangle) + O({\cal M}^3)\right]
(1+\delta_{\rm o})
\label{eq:deltag}
\end{equation}

\begin{figure}
\centering{
\includegraphics[width=0.49\textwidth]{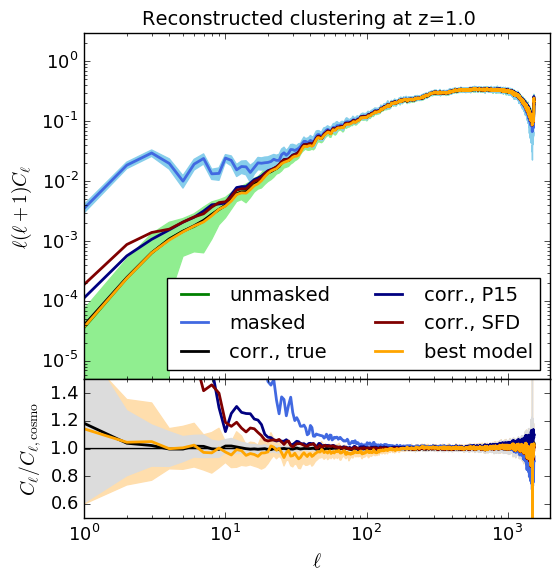}
\includegraphics[width=0.49\textwidth]{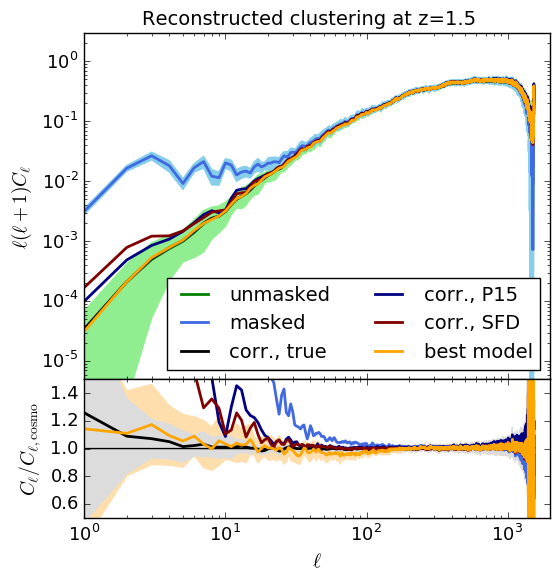}}
\centering{
\includegraphics[width=0.49\textwidth]{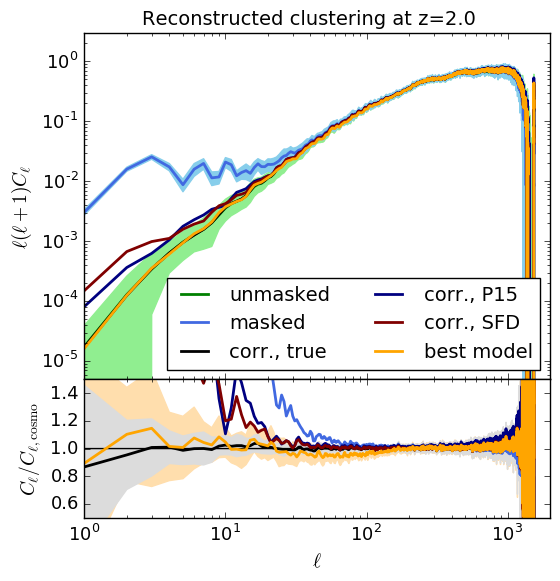}
\includegraphics[width=0.49\textwidth]{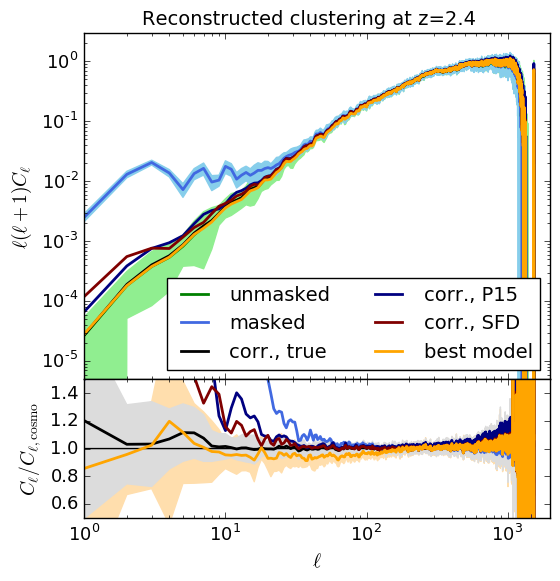}}
\caption{Reconstructed angular correlation of masked catalogs at
  redshifts $z=1.0$ (upper left), $1.5$ (upper right), $2.0$ (lower
  left) and $2.4$ (lower right). In the main panels the green and blue
  lines give the angular clustering of unmasked and masked catalogs,
  ligher colored areas giving the sample variance. Orange, black, dark
  blue and maroon lines give the average of the reconstructed angular
  correlation using respectively the best reconstruction, the true P13
  mask, the P15 and SFD masks. The lower panels give the residual with
  respect to the average true correlation of the unmasked mocks, the
  lighter coloured areas give the sample variance of the ratio of
  reconstructed and true clustering using the true mask (gray) and the
  best reconstruction (orange).}
\label{fig:reconstruction}
\end{figure}

Figure~\ref{fig:reconstruction} shows the angular power spectra of
galaxy density contrasts computed using equation~\ref{eq:deltag}, at
redshifts $z=1.0$, $1.5$, $2.0$ and $2.4$. The bright blue and green
lines show respectively the angular power spectrum of the (masked)
observed density contrast $\delta_{\rm o}$ and of the true galaxy
density contrast $\delta_{\rm g}$, the lighter-colored areas around
them give their sample variance (as in figure~\ref{fig:correlations}).
The recovery of the true galaxy power spectrum is performed using the
true P13 mask (black lines), the P15 (dark blue lines) and SFD (maroon
lines) masks, and the best reconstruction (orange lines). For each
panel, the lower sub-panel shows the ratio of the reconstructed power
spectrum over the one of the unmasked catalog, i.e. the quantity to
recover. Lighter-colored gray and orange areas give the sample
variances of the ratios obtained with the true mask and with the best
reconstruction.

\begin{figure}
\centering{
\includegraphics[width=0.75\textwidth]{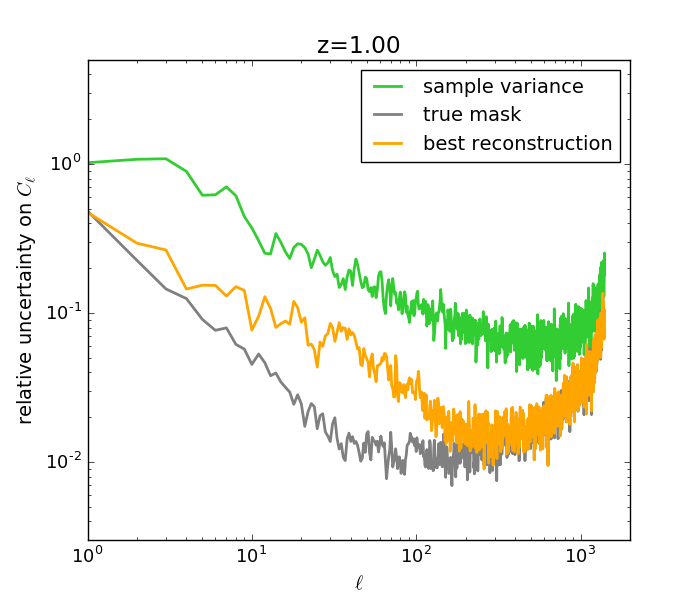}}
\caption{Ratio of sample variance over signal for the true angular
  power spectrum of galaxies (green line), for the ratio between
  reconstructed and true power spectrum using the true mask (gray
  line) and the best reconstruction (orange line).}
\label{fig:samplevariance}
\end{figure}

An important thing to notice is that the recovery of the true galaxy
density field is affected by some uncertainty even when the true mask
is used; this is due to the statistical nature of the correction
applied to the observed density contrast. To have a better
quantification of this, figure~\ref{fig:samplevariance} reports, for
the first redshift $z=1.0$ (other redshifts give very similar
results), the ratio of variance over signal for the cosmological
angular power spectrum and for the ratios of the two reconstructions
(with the true mask and with the best reconstruction). The uncertainty
in the reconstruction obtained with the true mask is always nearly a
factor of ten below sample variance, raising to higher values only at
high $\ell$'s, when the shot-noise-subtracted signal drops. The best
reconstruction carries a higher uncertainty by a factor of 3, that is
anyway lower by another factor of 3 than the sample variance; 
  however, it must be kept in mind that this bias is systematic, and
  cannot be decreased by averaging over several $\ell$'s.

As for averages, correcting with the true mask always gives an
unbiased reconstruction of the power spectrum, while correcting with
other masks gives results that significantly deviate from the true
clustering at low $\ell$'s; compared with the clustering of the masked
catalogs, correcting densities with the ``wrong'' mask gives an
improvement of only a factor of $\sim3-4$ in terms of the $\ell$ at
which clustering deviates significantly from the true solution.
Clearly deviations from true clustering are much less dramatic than
the option of not correcting at all, yet deviations at low $\ell$'s go
beyond the sample variance of the measurement.

The best reconstruction gives on average a remarkably unbiased
estimate down to very low $\ell$'s. Sample variance is important at
the largest angular scales, but always at the level of $\sim20$ \% and
below sample variance by a factor of 3. At $\ell=30-50$ we find a
systematic underestimation of $\sim5$ \%; it is plausible that this
bias could be removed by working on the combination of $E_{\rm av}$
and $E_{\rm sq}$ estimators, we leave this refinement to future work.

\subsection{Cross correlations as a blind test for the mask}
\label{sec:cross}

\begin{figure}
\centering{ \includegraphics[width=0.9\textwidth]{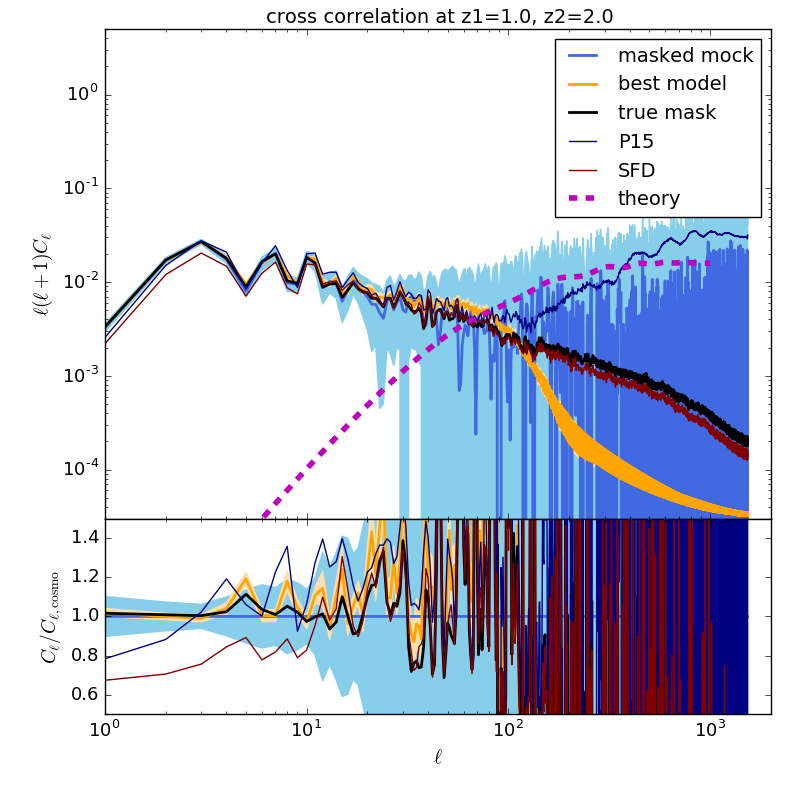}}
\caption{Angular cross-correlation of masked catalog for $z_1=1.0$ and
  $z_2=2.0$. The green and blue line gives the average measurement of
  the masked catalog, lighter blue areas gives its sample variance
  (16th and 84th percentiles). Orange, black, dark blue and maroon
  lines give the average of the predicted cross-correlation using
  respectively the best reconstruction, the true P13 mask, the P15 and SFD
  masks. The lower panels give the residual with respect to the
  average true cross correlation.}
\label{fig:cross}
\end{figure}

The cross-correlations of
$\delta_{{\rm o}1}\equiv\delta_{\rm o}(\boldsymbol\theta_1,z_1)$ and
$\delta_{{\rm o}2}\equiv\delta_{\rm o}(\boldsymbol\theta_2,z_2)$ for $z_1\ne z_2$
will depend only on the mask term, so it is possible to predict them,
given the mask reconstruction. Calling ${\cal M}_1={\cal
  M}(\boldsymbol\theta_1)$ and ${\cal M}_2={\cal
  M}(\boldsymbol\theta_2)$:

\begin{eqnarray}
  \langle\delta_{o1}\delta_{o2}\rangle &=&  F_1F_2\, 
  \langle({\cal M}_1-\langle{\cal M}\rangle)({\cal M}_2-\langle{\cal M}\rangle)\rangle
  +(F_1G_2 + F_2G_1)   \langle({\cal M}_1-\langle{\cal M}\rangle)({\cal M}^2_2-\langle{\cal M}^2\rangle)\rangle \nonumber\\
  &&+G_1 G_2\,  \langle({\cal M}_1^2-\langle{\cal M}^2\rangle)({\cal M}^2_2-\langle{\cal M}^2\rangle)\rangle\label{eq:crossmodel}\\
  F_i &=& \frac{S_{Ci}}{S_{Ai}}\left(1+ \frac{S_{Ci}}{S_{Ai}}\langle{\cal M}\rangle\right) \, , \ \ \ \ \ \ \ \ \ \ \
  G_i= \frac{S_{Bi}S_{Ci}^2}{S_{Ai}}\nonumber
\end{eqnarray}

\noindent
Figure~\ref{fig:cross} shows, as an example, the cross correlation of
observed density at redshifts $z_1=1.0$ and $z_2=2.0$. The blue line
gives the average measurement, the lighter blue area its sample
variance. Black, dark blue, maroon and orange lines give the predicted
cross correlation based respectively on the true mask, P15, SFD and
the best reconstruction, the last with its sample variance. While the black
line unsurprisingly gives the right level of correlation, the best
reconstruction (orange line) gives again a remarkably unbiased prediction of
this correlation, while the same is not true for P15 and SFD masks,
that show sizeable discrepancies in various $\ell$ ranges.

At the largest scales, while auto-correlations are dominated by the
foreground, for cross correlations this is the only contributing term.
In this paper we have constructed estimators of the non cosmological
term of observed density contrast, used them to reconstruct the mask,
and used the reconstructed mask to check that observed cross
correlations are correctly predicted. This logical progression could
have been reversed: we could have started from the measured cross
correlations of observed density contrast and use it to infer the
angular power spectrum of the mask. These two quantities are related
to the mask through equation~\ref{eq:crossmodel}, that however
includes correlations of the terms ${\cal M} - \langle{\cal M}\rangle$
and ${\cal M}^2 - \langle{\cal M}^2\rangle$. Even assuming knowledge
of the first two moments of the mask, it is not straightforward to
invert this relation to obtain $\langle{\cal M}_1{\cal M}_2\rangle$.
The presence of square terms of the mask in that equation is due to
the second-order expansion of the integral in
equation~\ref{eq:expanded}, and we know that a first-order expansion,
that would make the direct measurement of the angular power spectrum
of the mask possible, is not sufficiently accurate.

\begin{figure}
\centering{ \includegraphics[width=0.7\textwidth]{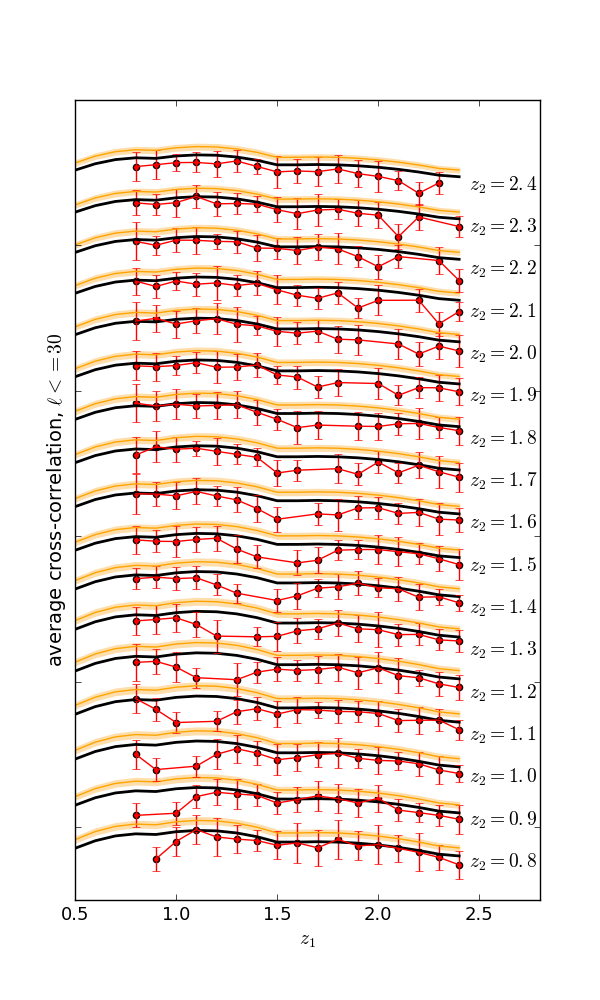}}
\caption{Average value of the cross correlation power spectrum
  $C_\ell$, for $\ell\le30$, of pairs of masked density fields at
  redshifts $z_1$ and $z_2$, as a function of $z_1$ and for all $z_2$.
  Points and curves are displaced vertically to distinguish the
  curves. Red points and errorbars give its average measurement (over
  the 20 mock catalogs) and sample variance (standard deviation).
  Black and orange lines give the predicted value using the true P13
  mask and the best reconstruction, results obtained with the
  ``wrong'' P15 and SFD masks are not shown for sake of clarity, but
  deviate significantly from the measurements. 
  }
\label{fig:crosses}
\end{figure}

To better quantify the level of agreement with which cross
correlations are recovered, we measure the average of
$\ell(\ell+1)C_\ell$ in the multipole range $\ell\le30$, where the
signal is strongest and sample variance is still limited
(figure~\ref{fig:cross}). Figure~\ref{fig:crosses} shows such average
cross correlation for all redshift pairs $z_1$ and $z_2$. Here the
dots show the average over 20 mocks and the errorbars give their
sample variance. Measurements are shown as a function of $z_1$, points
at fixed $z_2$ are displaced vertically to ease the comparison, so
vertical values are arbitrary. Points corresponding to
auto-correlations, with $z_1=z_2$, are not shown because they are
obviously affected by the cosmological signal. The black and orange
lines give the predictions obtained with the true mask and the best
reconstruction, for the latter we report its sample variance as a
lighter colored area. The prediction based on the true mask (black
line) follows well the data points; we computed a reduced $\chi^2$ to
test if the differences are significant, and found some systematics
due to nearby redshift bins, that show some degree of anti-correlation
of putative cosmological origin, and to the last bin, hinting to some
possible weakness in the mock construction near the edge of the light
cone. Excluding these bins (so limiting then the analysis to 91 bin
pairs), we found acceptable values of the reduced $\chi^2$, though the
prediction is biased high on average by 3\%. 

Being the cosmic correlation not exactly vanishing for nearby
  redshift bins, these could be removed from the estimate of $E_{\rm
    sq}$ to achieve a cleaner reconstruction; however, this advantadge
  would be counterbalanced by an increase of the residuals due to
  imperfect averaging out; given that cosmic correlation has in
  figure~\ref{fig:crosses} just an effect at the $2\sigma$ level,
  compared to sample variance, we prefer here to keep a larger number
  of bin pairs. We plan to deepen the effect of these correlations in
  future work.

Predictions of cross correlations based on the best reconstruction
show some mild discrepancy, that is quantified as a reduced $\chi^2$
of $\sim2$ for 91 measurements defined above. Predictions are biased
high by 10\%, that is at the level of $1\sigma$ but is systematic on
all measurements. As shown in the previous sections, the practical
effect of this bias is very small, but because cross correlations can
be measured, this bias can in principle be removed by further
calibrating the reconstruction scheme to best reproduce the average
cross correlations. One possibility could be to use this constrain to
measure at least one of the two moments of the mask; however the mask
enters equation~\ref{eq:crossmodel} through ${\cal M} - \langle{\cal
  M}\rangle$ and ${\cal M}^2 - \langle{\cal M}^2\rangle$, so cross
correlations are insensitive to the values of $\langle{\cal M}\rangle$
and $\langle{\cal M}^2\rangle$. We checked that it is possible to
obtain unbiased cross correlations by increasing $\langle{\cal
  M}\rangle$ by 0.04, leading to an almost doubled, unacceptable
value. Conversely, a multiplicative fudge factor applied to the
reconstructed mask of $0.97$ can make the predictions of the cross
correlations as (nearly) unbiased as those produced with the true
mask, with negligible impact on the pixel-by-pixel correlation of
figure~\ref{fig:bestmodel}.

We conclude this section stressing that, whatever the mask model or
reconstruction scheme is, cross correlations provide a blind test for
foreground removal. Suppose all known foreground corrections have been
applied to a deep galaxy sample, then one would expect cross
correlations to be consistent with a vanishing signal (quantified
through mocks), so a significantly non-vanishing value would show that
foreground removal is not complete. The angular power spectrum of the
cross correlations would immediately give clues on the nature of this
foreground, one could then calibrate the weights of the various
components to minimise the residual cross correlations, then apply the
reconstruction scheme presented here to model the missing residual,
and quantify its uncertainty using mock catalogs.

\section{Estimating $\langle {\cal M}\rangle$ and $\langle {\cal
    M}^2\rangle$}
\label{sec:moments}

\begin{figure}
\centering{ 
\includegraphics[width=0.49\textwidth]{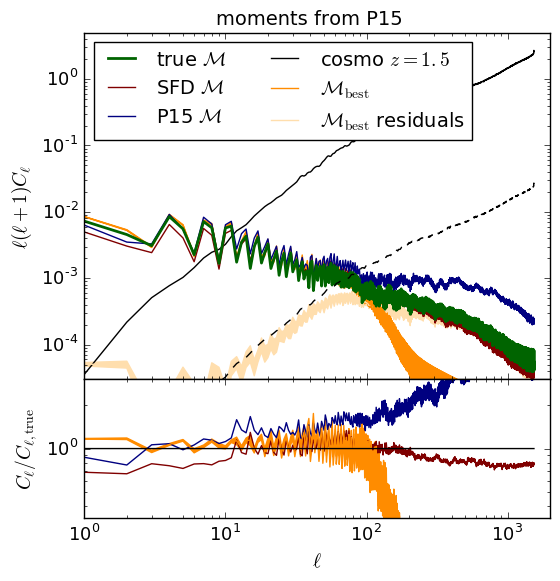}
\includegraphics[width=0.49\textwidth]{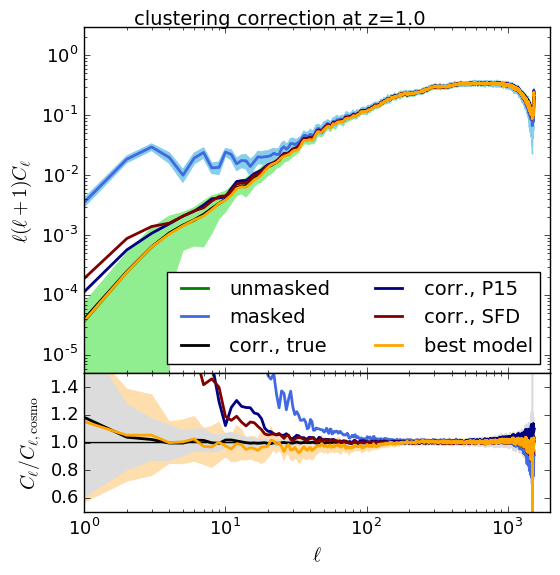}}
\centering{
\includegraphics[width=0.49\textwidth]{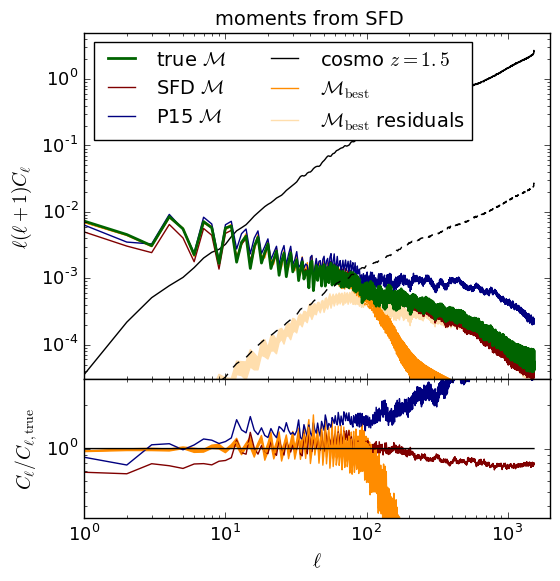}
\includegraphics[width=0.49\textwidth]{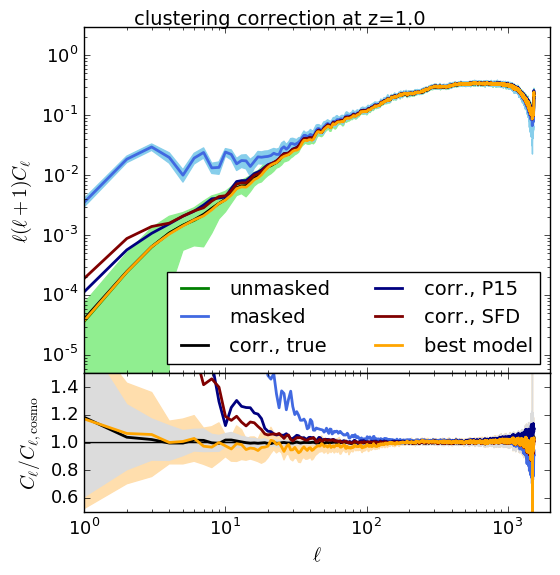}}
\caption{Angular clustering of best reconstruction for the mask (left panels)
  and reconstructed angular clustering of galaxies (right panels) for
  reconstructions that use $\langle {\cal M}\rangle$ and $\langle {\cal
    M}^2\rangle$ from P15 (upper panels) and SFD (lower panels).
  Symbols and colors are as in figures~\ref{fig:clbest} and
  \ref{fig:reconstruction}.}
\label{fig:momentsfrommaps}
\end{figure}

The results presented up to now rely on prior knowledge of the first
two moments of the mask, $\langle{\cal M}\rangle$ and $\langle{\cal
  M}^2\rangle$. We demonstrate in this section that it is possible to
recover this information either from external sources or internally,
without compromising the effectiveness of the method.

\begin{table}
\begin{center}
\begin{tabular}{l|cccccc}
\hline
Map & true & estimated & diff. & true & estimated & diff. \\
 & $\langle{\cal M}\rangle$ &  $\langle{\cal M}\rangle$ &  & 
$\langle{\cal M}^2\rangle$ & $\langle{\cal M}^2\rangle$  & \\
\hline
P13 & 0.0474 & 0.0453 & 4.5\% & 0.00356 & 0.00350 & 1.7\%\\
P15 & 0.0451 & 0.0457 & 1.1\% & 0.00382 & 0.00370 & 3.1\%\\
SFD & 0.0396 & 0.0417 & 5.5\% & 0.00277 & 0.00289 & 4.2\%\\
\hline
\end{tabular}
%% P13, <M> = 0.04745  -  <M^2> = 0.00356
%% P15, <M> = 0.04513  -  <M^2> = 0.00382
%% SFD, <M> = 0.03957  -  <M^2> = 0.00277
\end{center}
\caption{Values of the first two moments of the masks, and estimated
  values using the procedure described in this section.}
\label{tab:moments}
\end{table}

As a first step, we test the degradation of the results obtained using
the first two moments computed from a different reddening map, P15 or
SFD. When computed on the same sky area, the three maps give values of
$\langle{\cal M}\rangle$ that differ at most by 20\%, and values of
$\langle{\cal M}^2\rangle$ that differ at most by 38\%. These figures
are reported in table~\ref{tab:moments}; we can take their difference
as an order-of-magnitude indication of their measurement error.
Figure~\ref{fig:momentsfrommaps} shows the angular power spectrum of
the reconstructed mask (left figures, similar to
figure~\ref{fig:clbest}) and the reconstructed galaxy angular
clustering (right figures, similar to figure~\ref{fig:reconstruction})
at $z=1$, obtained using the values of the two moments from P15 or
SFD, in place of the true ones. Results show clearly that using the
``wrong'' values of the first two moments leads to nearly
indistinguishable results; the most remarkable difference is seen at
the largest scales, where the lower power of the SFD mask happens to
compensate the slight overestimate of the mask power spectrum
noticeable in figure~\ref{fig:crosses}, thus beating down the
residuals.

As mentioned above, the reason why results are insensitive to exact
values of the moments is visible in equation~\ref{eq:crossmodel}: the
mask enters this equation for the cross correlation through ${\cal
  M}-\langle{\cal M}\rangle$ and ${\cal M}^2-\langle{\cal
  M}^2\rangle$, plus a $\langle{\cal M}\rangle$ term in the $F_i$
coefficients, so a constant bias in the 1st moment of the mask has
only 2nd-order effects on the cross correlation, and consequently in
the ability to subtract the contamination from the cosmological
signal, while a bias in the 2nd order moment will have 3rd-order
effects that are neglected here.

\begin{figure}
\centering{  
\includegraphics[width=0.49\textwidth]{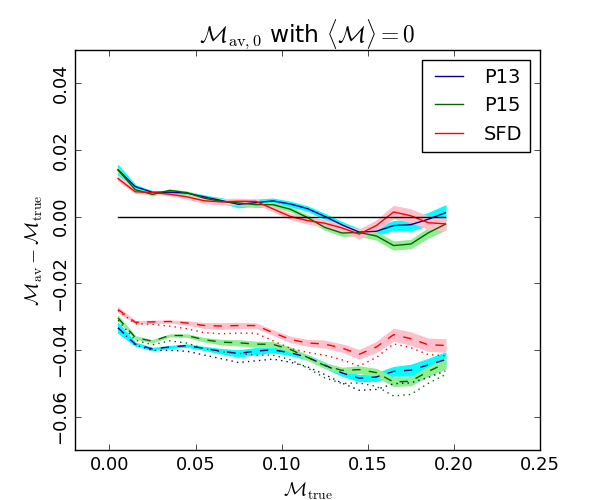}
\includegraphics[width=0.49\textwidth]{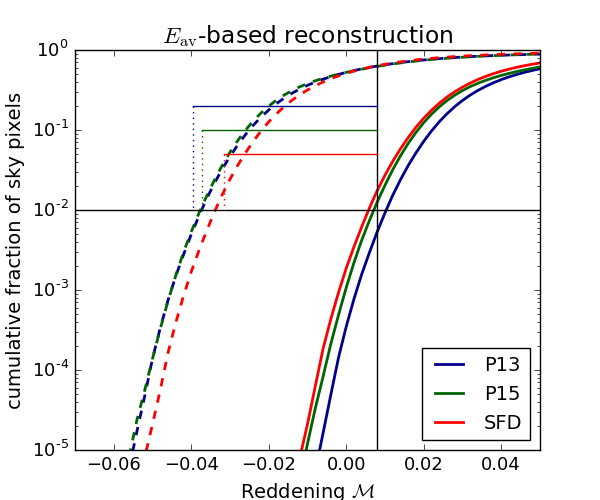}}
\caption{Left panel: average residuals of the ${\cal M}_{\rm av}$
  reconstruction (as the lower sub-panels of
  figure~\ref{fig:bestmodel} for mocks masked with the three reddening
  maps (P13, dark blue; P15, green; SFD, maroon), as a function of the
  true mask ${\cal M}_{\rm true}$ (different for the three cases).
  Continuous lines: $\langle{\cal M}\rangle$ and $\langle{\cal
    M}^2\rangle$ are computed from the true mask; the lighter-colored
  areas give the sample variance of the mean. Dotted lines: the same
  as the continuous line, shifted down by the corresponding value of
  $\langle{\cal M}\rangle$. Dashed lines: reconstruction computed with
  $\langle{\cal M}\rangle=0$ and $\langle{\cal M}^2\rangle=0$. Right
  panel: cumulative fraction of pixels as a function of reddening.
  Colors are as for the left panel, continuous lines give the true
  masks, dashed lines give results for ${\cal M}_{\rm av}$
  reconstructions with $\langle{\cal M}\rangle=0$ and $\langle{\cal
    M}^2\rangle=0$. The black horizontal line marks the fiducial
  $10^{-2}$ level chosen in the text, the vertical line gives the
  $0.008$ value of ${\cal M}_{1\%}$ (below which we have 1\% of sky
  pixels), the other segments mark the expected position of ${\cal
    M}_{1\%}$ for the $E_{\rm av}$-based reconstructions.}
\label{fig:get_aveM}
\end{figure}

We further show that it is possible to recover $\langle{\cal
  M}\rangle$ and $\langle{\cal M}^2\rangle$, to sufficient accuracy,
from an internal analysis of the survey data and some calibration on
the mock catalogs. To this aim we consider the $E_{\rm av}$-based
${\cal M}_{\rm av,0}$ reconstruction, applied to the survey assuming
zero values for the two moments. This is equivalent to applying the
$E_{\rm av}$ estimator, equation~\ref{eq:Eav}, directly to the
observed density $\delta_{\rm o}$ (equation~\ref{eq:deltaobs_m}) in
place of the rescaled $\delta_{\rm r}$
(equation~\ref{eq:deltaobs_resc}). At leading order in ${\cal M}$,
adequate for the least reddened sky pixels, it is easy to demonstrate
that $M_{\rm av,0}={\cal M}-\langle{\cal M}\rangle$, so a mask
reconstruction obtained assuming $\langle{\cal M}\rangle=0$ is
shifted, with respect to the true mask, exactly by $\langle{\cal
  M}\rangle$. This is shown on the left panel of
figure~\ref{fig:get_aveM}, where we report the average (and sample
variance) of ${\cal M}_{\rm av}-{\cal M}_{\rm true}$ obtained using
the mocks masked with the three P13, P15 and SFD masks, assuming exact
knowledge of the moments (the continous lines surrounded by lighter
colored areas) and zero moments (the dashed lines surrounded by
lighter colored areas). The quantities reported in this figure are
analogous to the residuals of ${\cal M}_{\rm av}$ shown in the upper
left panel of figure~\ref{fig:bestmodel} (without the errorbars used
to denote the scatter). It is worth highlighting that to produce this
figure we used mock catalogs masked with the three reddening maps, and
for each set of mocks we compare results with the corresponding true
map. The dotted lines here are the continuous lines shifted by the
corresponding value of $\langle{\cal M}\rangle$; they are shown to
demonstrate that the shift is very similar to the first moment of the
mask.

Unfortunately, the quantity reported in figure~\ref{fig:get_aveM} is
not observable.
We therefore identify a procedure to get a good estimate of
$\langle{\cal M}\rangle$. In the right panel of
figure~\ref{fig:get_aveM} we report the cumulative distribution of
reddening values obtained with ${\cal M}_{\rm av}$ applied to mocks
masked with the three reddening maps, both using the true
$\langle{\cal M}\rangle$ 
(continuous lines) and using $\langle{\cal M}\rangle=0$ (dashed
lines). These cumulative distributions are very robust to sample
variance, so we just show the average curve. To quantify the shift we
choose a {\em bona fide} fraction of $10^{-2}$, and quantify from
mocks that this fraction is obtained at ${\cal M}\simeq 0.008$ for the
three masks used. We define an estimate $\langle{\cal M}\rangle_{\rm
  est}$ of the first moment as the distance between 0.008 and the
${\cal M}_{1\%}$ value below which we find 1\% of sky pixels:

\begin{equation} \langle{\cal M}\rangle_{\rm est} = -{\cal
    M}_{1\%}+0.008\, . \label{eq:aveM} \end{equation}

\noindent
The obtained values are reported in table~\ref{tab:moments}, and are
accurate to within 5.5\% at worst. This demonstrates that such an
internal procedure, calibrated on mock catalogs, gives a potentially
more accurate estimate than using an external model for the map.

The estimate of $\langle{\cal M}^2\rangle$ is even less problematic,
because the reconstruction is not very sensitive to its exact value.
Once $\langle{\cal M}\rangle$ has been fixed, a very good estimate is
obtained by finding the best reconstruction ${\cal M}_{\rm best0}$
assuming a vanishing value of $\langle{\cal M}^2\rangle$, then using
the value of $\langle{\cal M}_{\rm best0}^2\rangle$. We find such
values to be biased high, for all three maps, by a constant $0.0007$,
that is broadly consistent with the idea that the final square average
is the sum of the true one and the squared scatter around the mean,
quantified here as $\sqrt{0.0007}=0.026$ mag. Using this
mock-calibrated fix, we obtain:

\begin{equation} \langle{\cal M}^2\rangle_{\rm est} = 
\langle{\cal M}_{best0}^2\rangle - 0.0007\, . \label{eq:sqaveM} \end{equation}

\noindent
The average values of this estimate for the three maps are reported in
the table; they are found to be within $\sim4.2$\% of the true value, a
very adequate level of accuracy.

\begin{figure}
\centering{ 
\includegraphics[width=0.49\textwidth]{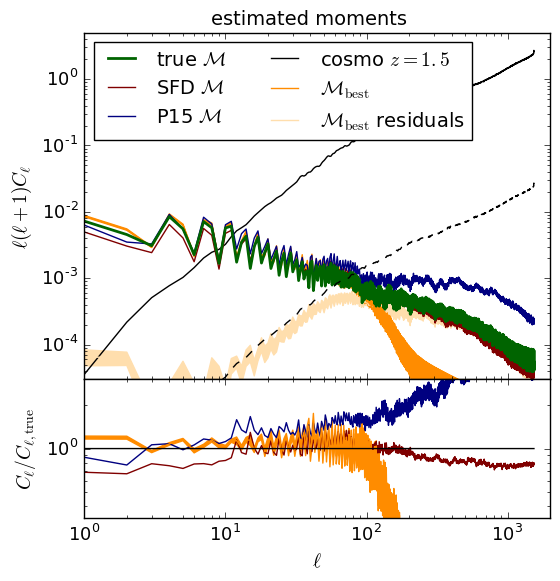}
\includegraphics[width=0.49\textwidth]{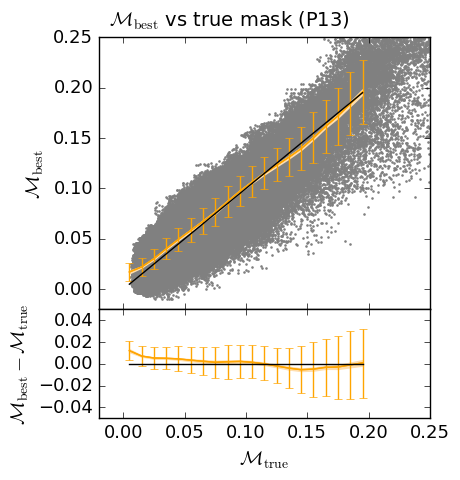}}
\centering{
\includegraphics[width=0.49\textwidth]{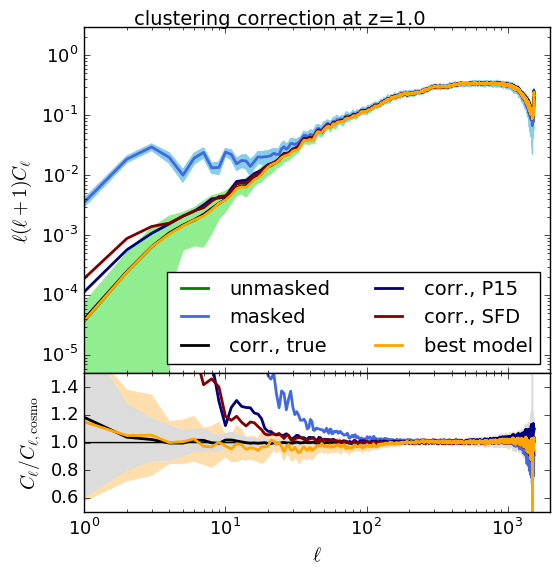}
\includegraphics[width=0.49\textwidth]{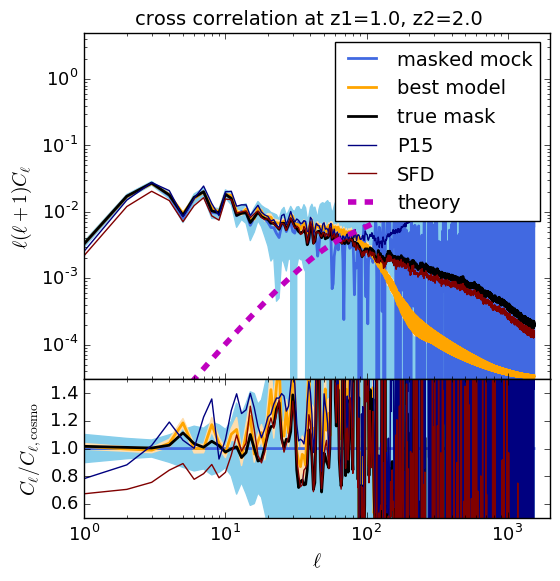}}
\caption{Performance of the best reconstruction obtained by estimating the
  first two moments of the mask from internal data. Upper left panel:
  angular power spectrum of the mask, as in figure~\ref{fig:clbest}.
  Upper right panel: pixel-by-pixel scatterplot of reconstructed and true
  masks, as in the lower right panel of figure~\ref{fig:bestmodel}.
  Lower left panel: reconstructed galaxy angular power spectrum at
  $z=1$, as in figure~\ref{fig:reconstruction}. Lower right panel:
  Angular cross correlation between $z=1$ and $z=2$, as in
  figure~\ref{fig:cross}. }
\label{fig:final}
\end{figure}

We are now in the position to present the best reconstruction of the
mask assuming no prior knowledge of its moment. We implement this
procedure: (i) the $E_{\rm av}$ estimator and ${\cal M}_{\rm av}$ are
computed assuming $\langle{\cal M}\rangle=\langle{\cal M}^2\rangle=0$;
(ii) equation~\ref{eq:aveM} is used to obtain an estimate of
$\langle{\cal M}\rangle$; (iii) the best reconstruction is constructed
assuming $\langle{\cal M}^2\rangle=0$; (iv) equation~\ref{eq:sqaveM}
is used to obtain an estimate of $\langle{\cal M}^2\rangle$; (v) the
best reconstruction is recomputed. One could further multiply the
result by $0.97$ to optimize the reproduction of the observed cross
correlations (Section~\ref{sec:cross}). Figure~\ref{fig:final} shows
the main results of the reconstruction: angular power spectrum of the
mask (as in figure~\ref{fig:clbest}), pixel-by-pixel scatterplot of
reconstructed and true masks (as in figure~\ref{fig:bestmodel}),
reconstructed correlation of the cosmic signal at $z=1$ (as in
figure~\ref{fig:reconstruction}) and predicted cross correlation
between $z=1$ and $z=2$ (as in figure~\ref{fig:cross}). None of the
results is found to suffer a degradation of accuracy.

\section{Testing reddening maps}
\label{sec:testing}

\begin{figure}
\centering{ 
\includegraphics[width=0.49\textwidth]{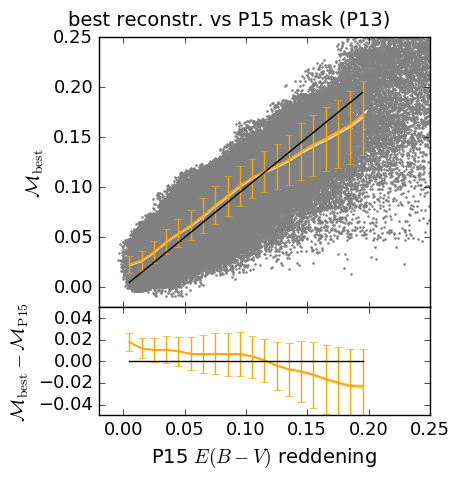}
\includegraphics[width=0.49\textwidth]{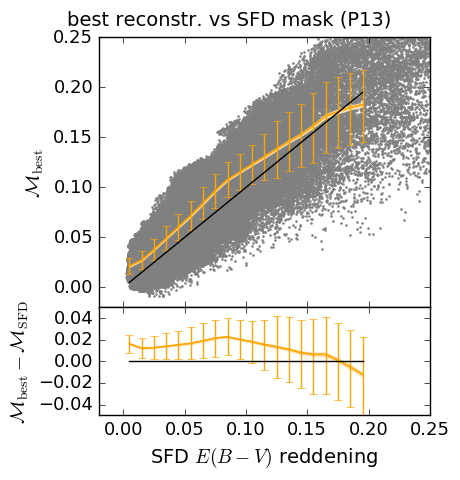}}
\centering{
\includegraphics[width=0.49\textwidth]{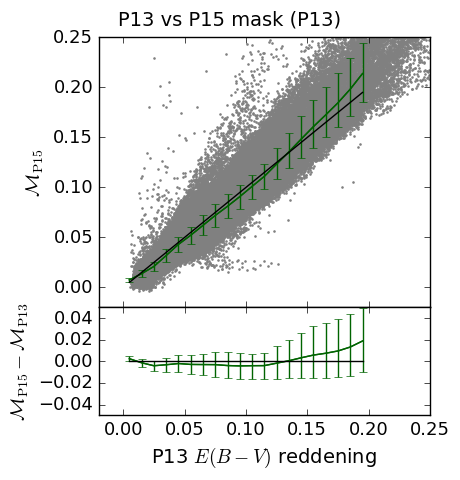}
\includegraphics[width=0.49\textwidth]{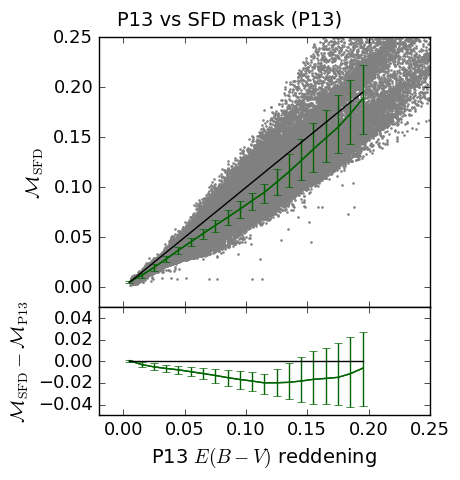}}
\caption{Upper panels: pixel-by-pixel scatterplots of the best reconstruction,
  obtained with mocks masked with the P13 reddening map, as a function
  of P15 or SFD reddening. Symbols and colors are as in
  figure~\ref{fig:bestmodel}. Lower panels: relation between P13, P15
  and SFD maps; here averages are of course computed for a single
  realization, so no sample variance is reported.}
\label{fig:testmaps}
\end{figure}

The results presented above demonstrate that the technique described
in this paper can recover the reddening map with a precision that is
comparable, and sometimes competitive, to the difference between
available reddening maps. The argument in principle can be reversed:
if extinction is the only foreground, reddening maps, obtained as
reconstructed foreground masks from deep and wide galaxy surveys, can
be used to constrain and test models for galaxy extinction, and thus
help in shedding light on the complex physics of dust in the solar
neighborhood. We have already shown evidence in this sense:
figure~\ref{fig:clbest} shows that the reconstructed power spectrum of
the mask is more compatible with the P13 power spectrum than with the
other two, P15 showing some power excess at $\ell\sim100$ and SFD some
lack of power at the largest angular scales. Figure~\ref{fig:cross}
shows that cross correlations are predicted to better accuracy by the
best reconstruction than by the P15 and SFD masks. If reported in
figure~\ref{fig:crosses}, predictions of P15 and SFD systematically
lie above (P15) and below (SFD) the measured values. It is important
to stress that this cross correlation test alone, that is completely
independent of the reconstruction procedure, would be enough to test
reddening maps and assess the one that best fits the data.

The same conclusion holds true at the pixel-by-pixel level.
Figure~\ref{fig:testmaps} shows the scatterplots of the best
reconstructions, as in figure~\ref{fig:bestmodel}, compared with the
P15 and SFD masks. For reference, the lower panels show the
scatterplot of P13 versus P15 and SFD. It is clear that, despite of
the higher level of scatter, the agreement of the best reconstruction
with the true P13 mask is better than the agreement with the other two
masks. This evidence alone would be enough to disfavour SFD, while P13
and P15 are nearly unbiased due to the very similar calibration on
quasars, and so are more difficult to distinguish with this test.

The weak point of this program is that it relies on the assumption,
true for our mock catalogs, that galactic extinction is the only
foreground that affects the survey. This will clearly not be the case
in general, and other foregrounds, for example contamination from
field stars, may be confused with extinction. However, different
foregrounds may be characterised by different $S_C$ scaling functions,
that quantify the redshift modulation of the impact of the foreground.
Two foregrounds that have proportional $S_C$ functions, e.g. a
redshift-independent constant, would not be separable; in the case of
$H\alpha$ emitters considered here, the effect of galactic extinction
is modulated in a peculiar way by the extinction curve $R(\lambda)$,
so its separation from other components is possible.

\begin{figure}
\centering{ \includegraphics[width=0.75\textwidth]{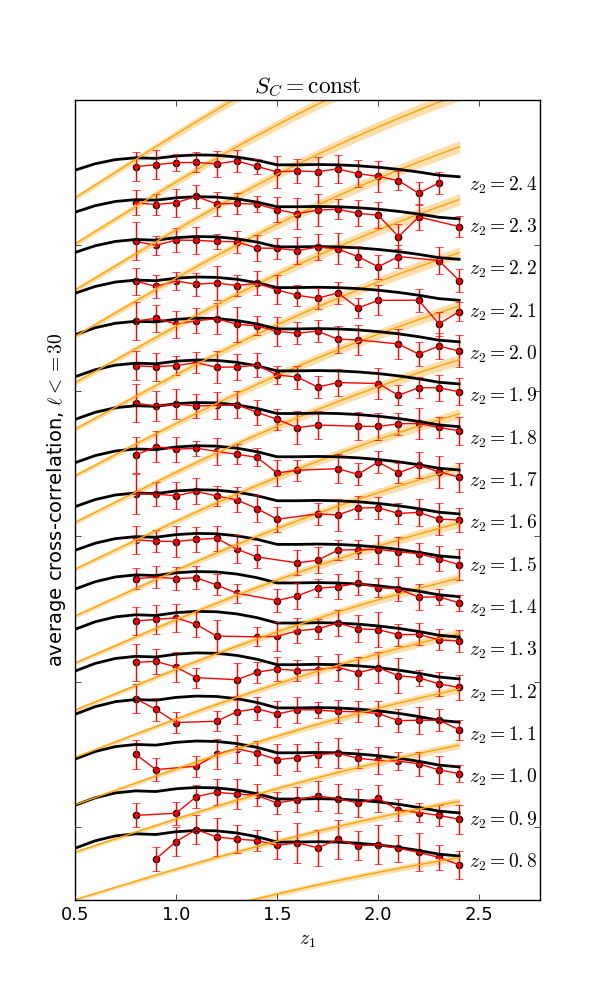}}
\caption{Average value of the cross correlation power spectrum
  $C_\ell$, for $\ell\le30$, of pairs of masked density fields at
  redshifts $z_1$ and $z_2$, as a function of $z_1$ and for all $z_2$.
  Symbols are as in figure~\ref{fig:crosses}, but best reconstruction
  predictions (orange lines) have been produced here by assuming a
  flat $S_C(z)$ function.}
\label{fig:flatC}
\end{figure}

The way to break the degeneracy between different foregrounds is to
fully exploit the ability to predict cross correlations at large
scales, e.g. as quantified in figure~\ref{fig:crosses}. As an example,
we show in figure~\ref{fig:flatC} the prediction of cross correlations
performed assuming a flat $S_C(z)$ function. While the angular maps
result very similar, the predicted redshift dependence of cross
correlations is in stark disagreement with observations. It would be
easy to assume a functional form for $S_C(z)$ (e.g. a polynomial) and
obtain the coefficients from fitting cross correlations; we plan to
test this procedure in the future.

\section{Conclusions}
\label{sec:conclusions}

In this paper we have proposed a blind method to determine the
properties of a foreground contamination that affects a deep galaxy
survey. As a prototypical foreground, we have considered Milky Way
extinction, and used three different extinction maps (Schlegel et
al's SFD and the two P13 and P15 Planck maps) available in the
literature. We have developed the method and tested its validity
using a set of 20 mock halo catalogs generated with the {\pin} code.
We have simulated an Euclid-like survey of $H\alpha$ emitters covering
$1/4$th of the sky in the redshift range $0.75<z<2.45$, by
abundance-matching the halo mass function with the luminosity function
of model 1 of Pozzetti; this way the impact of the
foreground is modulated in redshift by the extinction curve.

The method we propose is based on the fact that (i) foregrounds act in
modulating the flux limit of the galaxy survey, (ii) cross
correlations of galaxy density fields in different redshift bins have a
small cosmological contribution, dominated by lensing at $\ell>100$
and vanishing at larger angular scales. By expanding to second order
of flux modulation the observed number density of galaxies, we have
shown that the measured signal of cross correlations can be associated
to a foreground mask. We have identified two estimators, the average
density contrast $E_{\rm av}$ of a sky pixel along the line of sight
and the average $E_{\rm sq}$ of products of density contrasts at
different redshift bins in the same sky pixel, that average out (in an
ensemble-average sense) the cosmological signal, and single out a
contribution that depends only on the foreground mask. Assuming that
the first two moments of the foreground mask are known, it is possible
to relate these estimators to polynomials of the mask, that can be
easily inverted to reconstruct the visibility mask.

This reconstruction technique would be applied to a real survey
  as follows. (i) The luminosity function at the survey flux limit
  should be known with great detail from a deep field, so as to
  determine $S_A(z)$ and $S_B(z)$. (ii) Luminosity dependence of bias
  can be measured, up to a constant, by computing how the two-point
  correlation function (or power spectrum) changes when the sample is
  selected with an increasing flux limit. (iii) The $S_C(z)$ function
  must be first assumed, under some plausible hypothesis. (iv) The
  density contrast field is computed for each redshift bin. (v) The
  $E_{\rm av}$ and $E_{\rm sq}$ estimators are computed, the
  corresponding masks are obtained by inverting equations~\ref{eq:Eav}
  and \ref{eq:Esq} and the best reconstruction ${\cal M}_{\rm best}$
  is obtained through equation~\ref{eq:bestmodel}. (vi) The
  cross-correlation test is applied to the obtained mask to validate
  the assumed $S_C(z)$ function, if a good match is not found point
  (v) is iterated using some parameterization of this function, until
  a good fit of cross correlations is found. (vii) The mask thus
  obtained is used to produce a random catalog for clustering
  statistics. (viii) The mask can also be used to straightforwardly
  obtain the galaxy density contrast on the sky and the galaxy angular
  power spectrum using equation~\ref{eq:deltag}.

In the plausible case that different foregrounds, characterized
  by different $S_C(z)$ functions, are present at the same time, the
  method should be extended to take this higher complexity into
  account. Such extension is straightforward, and it would possibly
  benefit from the application of Bayesian inference methods based on
  forward modeling, as that proposed, e.g., by \cite{Porqueres2018}.

The reconstructed foreground mask reproduces very well, on relatively
large angular scales ($\ell<100$), the properties of the reddening map
used to mask the mock catalogs (P13), at a level that would make it
possible to distinguish it form P15 or SFD: extinction is recovered at
the pixel-by-pixel level with a bias that is below $0.01$ mag, and a
scatter ranging from $0.01$ to $0.03$ mag, slightly growing with
reddening. The reconstructed mask is smoothed in spherical harmonic
space at $\ell\sim100$, to remove the small-scale signal that is
dominated by the imperfect averaging out of the cosmological signal,
due both to the limited number of redshift bins and to the use of
volume averages in place of ensemble averages. The so-obtained
reconstruction of the mask can be used to remove the spurious power
from the angular correlation function. We showed that this removal
gives unbiased results up to $\ell$ of a few, where the cosmological
signal is two orders of magnitude below the contamination, with a
statistical uncertainty of $\sim20\%$, a  (systematic) factor of 3 below the sample
variance of the observable. This technique can then have an impact on
constraints on primordial non-gaussianities, whose measurement is
limited by sample variance, thoug constraints can be improved by
adopting multi-tracer techniques~\cite{Seljak:2008xr} to beat cosmic
variance. These techniques allow to extract information also from the
largest scales probed by the surveys. Indeed, as shown in
Ref.~\cite{Alonso:2015sfa}, it is crucial to being able to recover
information from scales larger than $\ell_{\rm min} \lesssim 10$ to
provide constraints of the order $\sigma \left( f_{\rm nl} \right)
\sim \mathcal{O} \left( 1 \right)$, which will be able to discriminate
between different inflationary models. We leave to further work an
assessment on how well our approach will help in this field.

One could take from this analysis the main message and use cross
correlations of different redshift bins as a powerful blind diagnostic
of the presence of residual foregrounds, after having removed all
known sources. This would show both residual contaminations and
possible effects of ``unknown unknowns'', while the redshift
dependence of cross correlations would help in modeling and
constraining the residual foregrounds. Or one could take a more
aggressive path, and apply this method to raw data, modeling the
resulting mask as the sum of plausible elements with their $S_C(z)$
functions. Once the components are identified, some sophisticated
statistical approach, analogous to component separation for the CMB,
would allow a complete characterization of foregrounds. 

It is worth to list here and discuss the assumptions that have been
done to obtain this result. 

The effect of foregrounds has been assumed to be the modulation of
survey depth on the sky. This is fine for galaxy clustering, but other
observables will need different formalizations; for instance, fiber
collision would impact not simply at the flux limit but in a (known)
range of fluxes, and would mostly add spurious signal to small scales;
conversely, 21 cm intensity mapping would sample the whole luminosity
function, so the noise would be purely additive.

The luminosity function has been expanded to second order in the
modulation of flux limit. We have found a first-order expansion to be
inaccurate already at reddening values of $\sim0.05$, while a
second-order expansion is fine at least to $E(B-V)\sim0.1$, where most
sky pixels lie in typical extragalactic surveys. An eventual expansion
to higher order would further complicate the algebra and would require
knowledge of ${\cal M}^3$ and of the second derivative of the
luminosity function and bias, but would be feasible.

The shape of the luminosity function has been assumed to be universal.
We know that this is not true (see e.g. \cite{Zucca2009}, or
  \cite{Malavasi2017} for the environment dependency of the stellar
  mass function), and the way this universality is violated depends
crucially on density and in general on galaxy selection. However, this
is a general problem, any correction of galaxy number counts to
absorb, say, a known modulation of flux limit will rely on knowledge
of the slope of the luminosity function. The only way to tackle this
issue is to make many simulations and tests with plausible variations
of $\Phi(L)$.

Galaxy bias has been treated at a linear level. More importantly, the
luminosity dependence of (linear) bias has been taken into account in
the analytic development, but the tests performed in this paper have
been done by removing this dependence by shuffling halo masses before
abundance-matching them with $H\alpha$ galaxies. While linear bias is
considered adequate at the very large scales where a foreground like
the Milky Way has an impact, the removal of luminosity dependence of
bias has been done in this paper only for sake of simplicity. This
assumption does not influence the mask reconstruction, because bias
enters only the cosmological term $T_{\rm c}\delta$ of $\delta_{\rm
  o}$ in equation~\ref{eq:deltaobs} that is averaged out by the
estimators. It would only complicate the comparison with linear theory
predictions and the recovery of true clustering.

The luminosity function and the extinction curve have been assumed to
be know with arbitrarily good accuracy. In other words, the $S_A(z)$,
$S_B(z)$ and $S_C(z)$ functions have been assumed to be known without
error. The uncertainty on the luminosity function for a survey with
millions of galaxies is expected to be so small that its error is
  going to be negligible and violations of the universality of
  $\Phi(L)$ are likely to be much more serious. Conversely, as
commented in Section~\ref{sec:cross}, one could use cross correlations
to measure $S_C(z)$ (and the extinction curve in case the foreground
is galactic extinction). Results obtained assuming a flat $S_C(z)$
function have been shown to provide completely wrong cross
correlations but a very similar angular mask ${\cal M}$, so the
uncertainty in $S_C(z)$ is likely to have little impact on the
reconstructed foreground mask.

Foregrounds have been assumed to be uncorrelated with the cosmological
signal, i.e. $\delta({\mathbf x})$ and $\delta L(\boldsymbol\theta)$
are statistically independent. This is true in an ensemble average
sense but not for volume averages, and this contributes to the
residuals. Also, this is not necessary true in some cases. For
instance, if magnitudes are corrected using an extinction map based on
FIR dust emission, fluctuations of the Cosmic Infrared Background
(CIB) could be present in this correction in case dust emission is not
perfectly separated from the CIB. This worry has been taken seriously,
e.g.~by the Planck collaboration in the presentation paper of the P13
map. However, the power spectrum of CIB fluctuations should parallel
the cosmic signal, that is steeply rising (see
figure~\ref{fig:clmask}), while the angular power of the reddening
masks we have used show a declining spectrum, hinting to a low
contamination of this type. A more complicated case arises when the
contamination correlates with source density, like for fiber collision
or overlapping spectra in slitless spectroscopy. In this case the
averaging out of cosmic signal is not obvious, and tests would be
needed to estimate the impact of this correlation and model its
behaviour.

Because light cone effects are not present in the mocks that we have
produced, the role of lensing has only been estimated to affect a
different range of scales with respect to extinction.
Luminosity-dependent bias would further boost this effect at high
redshift, creating a potential overlap that should be modeled. The
redshift dependence of the impact of lensing can be computed
theoretically, so it would be easy to model its impact on the cross
correlations and subtract it out. However, the accuracy of this
subtraction must be assessed.

The main points of strength of the method we are proposing are the
following:

(i) It requires a very limited set of assumptions on the foreground.
The first two moments of the mask, that are needed to build the best
reconstruction, can be estimated from the data with some calibration
on the mock catalogs, and the subtraction of the cosmic signal depends
very mildly on their exact value. Knowledge of the $S_C(z)$ function,
that modulates in redshift the impact of the angular mask, is crucial,
but it can be recovered from the redshift evolution of the cross
correlations.

(ii) Because the mask is reconstructed from the same survey that needs
to be corrected, it is accurate exactly where it needs to be. For
instance, for the Milky Way extinction case considered here the
foreground has most power on large angular scales, so the
reconstructed mask is dominated by imperfect foreground removal at
$\ell>100$. But, thanks to the the available number of independent
redshift bins, the level of contamination is below 1\% of the
cosmological signal; the mask cannot be recovered only because it has
such a low angular power that it is not going to affect the survey. A
foreground with a flatter or mildly raising angular power spectrum
would be measurable at much higher $\ell$'s.

(iii) It is straightforward to compute the uncertainty with which the
foreground mask is recovered, by applying this method to a large set
of simulated catalogs. The covariance matrix of clustering statistics
is usually computed using mock catalogs, so one would just need to
apply the method to the same set of mocks used for the cosmological
covariance, masking them with some model mask that augments the best
reconstruction obtained from data with high-$\ell$ power. Such a
covariance matrix would then account for the residual correlation of
the reconstructed mask with the cosmological signal, due to its
imperfect averaging out by the $E_{\rm av}$ and $E_{\rm sq}$
estimators. This is a crucial point of merit: one could get from
figure~\ref{fig:testmaps} that the uncertainty in the recovery of the
mask, at the pixel-by-pixel level, is larger than the difference
between the available maps, but it is possible to fully characterise
this uncertainty and its contributions to the covariance matrices of
any statistics for which the cosmic covariance is computed with mock
catalogs. The same is not true for reddening maps recovered from FIR
observations: an unknown gradient in dust composition or properties
across the sky would give highly correlated errors that would be
almost impossible to characterize.

(iv) This method can be applied to photometric redshift catalogs.
Clearly, nearby redshift bins would be contaminated by the redshift
uncertainty, to a level that would be easy to detect from cross
correlation (see figure~\ref{fig:crosses}), so the number of usable
redshift bin pairs would be lower; but it would be easy to get deep
catalogs with much lower shot noise. One could also strengthen the
result using Lyman-break galaxies or quasars at much higher redshift.
In principle, it is straightforward to add information from
complementary surveys, even though the control of uncertainties would
be complicated by the need to have mocks that model several types of
objects at the same time.

(v) Finally, cross correlations could be a powerful diagnostic for
catastrophic redshift errors. Figure~\ref{fig:crosses} would show
peaks at specific redshift intervals that correspond to misinterpreted
lines, if this contamination is significant.

\section*{Acknowledgements}

This paper has benefited of discussions and the stimulating
environment of the Euclid Consortium, which is warmly acknowledged.
P.M. acknowledges stimulating discussions with Luigi Guzzo, Marta
Spinelli, Ben Granett and Francisco Villaescusa-Navarro. We are
grateful to Healpy developers for having provided such a useful and
coder-friendly tool. {\pin} has been run on the GALILEO@CINECA
supercomputer with computing time budget from INAF CHIPP and the
agreement UNITS-CINECA. P.M. and E.S. acknowledge support from PRIN
MIUR 2015 {\em Cosmology and Fundamental Physics: illuminating the
  Dark Universe with Euclid} and from INFN InDark research project.
E.D. is supported by the Swiss National Science Foundation (No.
171494).

\bibliographystyle{JHEP}
\bibliography{bibliography} 

\end{document}